\documentclass[%
 reprint,
nobibnotes,
 amsmath,amssymb,
 aps,
prl,
]{revtex4-2}

\usepackage{graphicx}
\usepackage{dcolumn}
\usepackage{bm}
\usepackage{bbm}

\usepackage{color}
\usepackage{hyperref}
\usepackage{ulem}

\begin{document}


\title{Altermagnetic Instabilities from Quantum Geometry}

\author{Niclas Heinsdorf*}
\affiliation{Max-Planck-Institut für Festkörperforschung, Heisenbergstrasse 1, D-70569 Stuttgart, Germany}
\affiliation{Department of Physics and Astronomy \& Stewart Blusson Quantum Matter Institute,
University of British Columbia, Vancouver BC, Canada V6T 1Z4}
\email{heinsdorf@fkf.mpg.de}

\date{\today}

\begin{abstract}\noindent
Altermagnets are a newly identified type of collinear anti-ferromagnetism with vanishing net magnetic moment, characterized by lifted Kramers' degeneracy in parts of the Brillouin zone. Their time-reversal symmetry broken band structure has been observed experimentally and is theoretically well-understood. On the contrary, altermagnetic fluctuations and the formation of the corresponding instabilities remains largely unexplored. We establish a correspondence between the
quantum metric of normal and the altermagnetic spin-splitting of ordered phases. We analytically derive a criterion for the formation of instabilities and show that the quantum metric favors altermagnetism. We recover the expression for conventional q=0 instabilities where the spin-splitting terms of the normal state model are locally absent. As an example, we construct an effective model of MnTe and illustrate the relationship between quantum geometry and altermagnetic fluctuations by explicitly computing the quantum metric and the generalized magnetic susceptibility. 
\end{abstract}

\maketitle

\section{Introduction} Recognizing the importance of the geometrical aspects of the manifold that is formed by a set of Bloch states has revolutionized condensed matter physics and material science\cite{berry1984quantal,resta_polarization,review_berry}. Understanding its relation to band structure and crystallography has enabled a systematic search for topological materials on a large scale\cite{jan-slager,Bradlyn2017,Vergniory2019,Tang2019,Zhang2019}. Nowadays, there are numerous generalizations that aim to reconcile this concept with many-body physics including interacting topological quantum chemistry\cite{mikel1,dominik,marti_itqc,mikel2,Herzog-Arbeitman2024}, symmetry-protected topological phases\cite{pollmann1,pollmann2,chen1,chen2,spt_review}, many-body Chern numbers\cite{mbcn,mbc_nointegration}, band topology of spin excitations\cite{hosho1,hosho2,topmagnonreview,heinsdorf2023stablebosonictopologicaledge}, higher Berry curvature\cite{kapustin,flowhigher,ken,ophelia1,ophelia2}. In principle, topological invariants do not depend on local terms, but in practice, conventional band topological invariants are computed as an integral of individual geometric contributions over the full parameter space. Quantum geometry -- including both the quantum metric and Berry curvature\cite{essay,cheng2013quantumgeometrictensorfubinistudy} -- arises from the non-zero overlap of well-defined eigenstates that are close in parameter space, or alternatively is encoded in the rates of changes of band projectors and their associated (generalized) Bloch vectors\cite{berry1989quantum,wilczek1989geometric,quantum_metric_Nband}. In any case, both points of view require an effective description in terms of quasi-particles or ``bands", which is generally obfuscated by strong interactions. 

All the more surprising, the quantum metric has lately been proven to be a powerful tool to make predictions about what are innately interacting problems. In experiment, it can be measured directly\cite{measure1,measure2,measure3,measure4,measure5,measure6}, and on the theoretical side it has been used to establish bounds on the resistivity, capacitance and Drude weights of (correlated) systems\cite{bound_johannes,Komissarov2024,bounds_drude}, helped to explain flat-band superconductivity\cite{sf_weight1,sf_weight2}, (anomalous) fractional Chern insulators\cite{vortexability1,vortexability2}, Landau level spreading in flat bands\cite{LLspread}, and was shown to enhance Fulde-Ferrell-Larkin-Ovchinnikov, anapole and spin-triplet superconductivity\cite{taisei_FFLO,taisei_anapole,taisei}. In this work, we argue that the quantum metric is a key feature of the normal-state band structures of altermagnets, and then explicitly show that it favors the formation of altermagnetic instabilities by analyzing their fluctuations. 

Needless to say, magnetism is extremely relevant to modern-day technology, and altermagnetism -- which has been observed experimentally for the first time only earlier this year\cite{mnte_prl,mnte_nature} -- holds the potential to be a key ingredient for many  spintronic or other future device applications\cite{hayami1,hayami2,libor1,libor2,am_review2,am_review1,igor_review,spintronic1,spintronic2,spintronic3,jenn_am_sc,am_andreev}. The group theoretical classification of altermagnets is ongoing, but nevertheless well understood\cite{ssgs1,ssgs2,ssgs3,ssgs4,alberto,ssgs5,parshukov2024topologicalresponsesgappedweyl}. In contrast, the nature and intrinsic structure of altermagnetic fluctuations and instabilities continue to be mostly uncharted\cite{roig2024minimal}. Here, we elucidate its intricate connection to quantum geometry. 

\section{Quantum Metric in Altermagnets}
Altermagnetism is a special type of collinear, compensated anti-ferromagnetic order with lifted Kramers' degeneracy in parts of the Brillouin zone (BZ). There are essentially two ways of thinking about it: (i) We can understand altermagnetism as a ``less symmetric" version of anti-ferromagnetic order. In a conventional anti-ferromagnet the two magnetic sublattices are related by inversion or a fractional translation symmetry (w.r.t. the magnetic unit cell), which leads to a global band degeneracy at any momentum. The sublattices in an altermagnet are mapped onto each other by an (improper) rotation $A$, resulting in degeneracies only at momenta that are invariant under the action of any point group symmetry containing $A$\cite{am_review2}. (ii) Alternatively, we can recognize that altermagnets do require some minimal symmetry (said rotation) that relates their spin-up and -down sublattices, because otherwise they would hardly be (perfectly) collinear and compensated. From this angle, altermagnets are not spin-split anti-ferromagnets, but rather platforms with extended band degeneracies as enforced by crystal symmetry -- just like nodal line\cite{nl1,nl2,mo_larhge}, loop\cite{drumhead1,drumhead2}, chain\cite{tomas_chain,irf4_chain} or plane\cite{tetragonal_np,orthorhombic_np,fundamental_np,mnsi_np,cosi_np,conbs_np} semimetals, and indeed, the Berry phase around a loop enclosing an altermagnetic nodal line is $\pi$-quantized and Hall, as well as non-linear responses can be generated through inclusion of different symmetry-breaking mass terms\cite{parshukov2024topologicalresponsesgappedweyl,cano_am_transport_nonlinear}.

Regardless of the viewpoint, the classification of altermagnets is essentially a group theoretical problem, and just by analyzing the space group symmetry of the normal state (above N\'eel temperature $T_N$), we can distinguish altermagnets from anti-ferromagnets even though the magnetic order parameter is exactly the same. However, group theory cannot explain why an instability towards altermagnetism is preferred over, for example, a ferromagnetic instability. This is where perspective (ii) is helpful, because it highlights the restrictions imposed by symmetry on the normal state band structure which, in turn, determine the bare susceptibility. These symmetries enforce extended band degeneracies, which suggests that interband effects are important in forming altermagnetic instabilities. This hypothesis is supported by the findings in Ref.\ \cite{roig2024minimal}. Therein, the authors investigate generic (two-band) altermagnetic tight-binding models, and compute non-interacting susceptibilities given by 
\begin{align}\label{eq:sus_dens_dens}
    \chi(i\Omega_n, \mathbf{q}) = \int_0^\beta d\tau e^{i\Omega_n \tau} \langle T_\tau  \tilde{n}(\tau, \mathbf{q}) \tilde{n}(0, \mathbf{-q})\rangle,
\end{align}
where $\tilde{n}$ is a (modified) density operator $\tilde{n}(\tau, \mathbf{q}) = \frac{1}{N}\sum_\mathbf{k}\mathbf{c}^\dagger(\tau,\mathbf{k})\tau_{\mathrm{M}}\mathbf{c}(\tau,\mathbf{k} + \mathbf{q})$. The inserted operator $\tau_{\mathrm{M}}$ lives in the space of the bipartite, magnetic sublattice and toggles between the conventional density-density susceptibility for $\tau_{\mathrm{M}}=\mathbbm{1}_{2\times 2}$ on the one, and the altermagnetic susceptibility $\chi^{\mathrm{AM}}$ for $\tau_{\mathrm{M}}=\tau_z$ on the other hand -- with $\tau_z$ being the third Pauli matrix. Because $\tau_z$ is not diagonal in the band basis, it is clear immediately that interband contributions must enter, and for the models considered here, it can easily be shown that an AM instability is preferred as long as there is finite intersite coupling\cite{roig2024minimal}.

Another piece of evidence for the importance of interband effects in these models is their quantum geometry. Because the normal state band structure of altermagnets is mostly characterized by pair-wise degeneracies (and because contributions of more distant band-pairs fall off quickly), it can be generically described by two-band models of the form\cite{roig2024minimal}
\begin{align}\label{eq:normalstate_hamiltonian}
    \mathcal{H}_0(\mathbf{k}) = \varepsilon_0(\mathbf{k})\mathbbm{1}_{2\times 2} + \left( t_x(\mathbf{k}), 0, t_z(\mathbf{k})\right) \cdot \boldsymbol{\tau}
\end{align}
with $\boldsymbol{\tau} = (\tau_x, \tau_y, \tau_z)^\top$ and the sublattice-independent dispersion $\varepsilon_0$. The inter- and intra-sublattice hopping coefficients $t_x$ and $t_z$ transform even and odd under any point group symmetry that exchanges the two sublattices respectively. At sufficiently low temperatures, altermagnets order magnetically. The corresponding N\'eel order parameter $\Delta$ describes collinear, but fully compensated order that enters on the level of the effective Hamiltonian as $\mathcal{H}_{\mathrm{N}} = \Delta\tau_z\sigma_z$. In the absence of spin-orbit coupling (SOC), the lattice and spin degrees of freedom are completely decoupled, and we can choose w.l.o.g. the direction of magnetic moment to be along the $z$-direction. We want to stress that $\mathcal{H}_{\mathrm{N}}$ looks just like for a conventional anti-ferromagnet. In fact, if $t_z$ was zero, the spin-splitting, which is a defining characteristic of an altermagnet, would be absent -- even for finite $\Delta$. 

Our work is largely inspired by the fact that this is reflected in the quantum metric of the normal state Hamiltonian $\mathcal{H}_0$ already. In terms of Bloch vectors, the quantum metric can be expressed as\cite{quantum_metric_Nband}
\begin{align}\label{eq:quantum_metric_bloch}
    g_{\pm,\mu\nu}(\mathbf{k}) = \frac{1}{4}\partial_{k_\mu}\mathbf{b}_\pm(\mathbf{k})\partial_{k_\nu}\mathbf{b}_\pm(\mathbf{k}).
\end{align}
The Bloch vectors of the two-band model are given by $\mathbf{b}_\pm(\mathbf{k}) = \pm\left( t_x(\mathbf{k}), 0, t_z(\mathbf{k})\right) / \sqrt{t_x^2(\mathbf{k}) + t_z^2(\mathbf{k})}$, or in terms of the Bloch angle $\theta$ as 
\begin{align}
    \mathbf{b}_\pm(\mathbf{k}) = \pm\left( \sin \theta(\mathbf{k}), 0, \cos \theta(\mathbf{k})\right).
\end{align}
For non-zero $t_z$, the quantum metric is generally finite, but for $t_z = 0$ (the anti-ferromagnetic case) the Bloch vector is no longer momentum-dependent, which results in zero quantum metric at all $\mathbf{k}$. Notably, $t_x = 0$ leads to zero quantum metric as well, and while it does not affect the spin-splitting, Ref.\ \cite{roig2024minimal} found that when it is zero, the AM and FM susceptibilities are degenerate. 

We argued that the quantum metric is not only a measure by which altermagnets can be distinguished from conventional anti-ferromagnets (without even knowing the exact space group symmetries), but moreover that it can be indicative of whether or not an altermagnetic instability is favored. Pursuing this intuition, we analyze the altermagnetic fluctuations of general altermagnetic models in their normal state and explicitly identify the quantum metric in the resulting expression.

\section{Main Results} We define the generalized magnetic susceptibility (GMS) $\chi^{0:\mu\nu}_\mathrm{m}$ which is an extension of the generalized electric susceptibility (GES) introduced in Ref.\ \cite{taisei} as $\chi^{0:\mu\nu}_\mathrm{e} = \lim_{\mathbf{q}\rightarrow 0} \partial_{q_\mu}\partial_{q_\nu}\chi(0, \mathbf{q})$ (with $\tau_{\mathrm{M}}=\mathbbm{1}_{2\times 2}$). In insulators, the GES gives corrections to the equilibrium charge density, and in metals (at finite temperature) can be related to the electric quadrupole moment\cite{taisei,yanase_meaning_ges}. $\tau_{\mathrm{M}}$ switches between $\chi^{0:\mu\nu}_\mathrm{e}$ and $\chi^{0:\mu\nu}_\mathrm{m}$, which quantify the role of ferro- and altermangetic fluctuations respectively. For our purposes, it is sufficient to think of it as a second-derivative test: A peak of $\chi^{\mathrm{AM}}$ at $\mathbf{q}=0$ corresponds to altermagnetic fluctuations, so a criterion for the formation of an instability is $\chi^{0:\mu\nu}_\mathrm{m} < 0$, that is negative curvature at that point. Dissecting $\chi^{0:\mu\nu}_\mathrm{m}$ into positive and negative contributions, we make statements about what terms are favourable or detrimental to altermagnetism. The detailed derivation of $\chi^{0:\mu\nu}_\mathrm{m}$ can be found in the supplemental material\cite{supp}. For altermagnetic models of the form Eq.\ \ref{eq:normalstate_hamiltonian} it simplifies drastically, and we write the (diagonal part of the) GMS as $\chi^{0:\mu\mu}_\mathrm{m:tot} = \chi^{0:\mu\mu}_\mathrm{m:mass} + \chi^{0:\mu\mu}_\mathrm{m:velo} + \chi^{0:\mu\mu}_\mathrm{m:geom}$ with
\begin{align}\label{eq:chi_mass}
    \chi^{0:\mu\mu}_\mathrm{m:mass} = \frac{1}{6}\int_{\mathrm{BZ}}\cos^2\theta(\mathbf{k})\sum_{n=1}^{2} f^{(3)}[E_n(\mathbf{k})]v^2_{n,\mu}(\mathbf{k}) 
\end{align}
\begin{align}\label{eq:chi_velo}
    \chi^{0:\mu\mu}_\mathrm{m:velo} = 2\int_{\mathrm{BZ}}\sin^2\theta(\mathbf{k}) F[E_{1}(\mathbf{k}), E_{2}(\mathbf{k})]\frac{v_{1,\mu}(\mathbf{k})v_{2,\mu}(\mathbf{k})}{\left(E_1(\mathbf{k})-E_2(\mathbf{k})\right)^2}
\end{align}
\begin{align}\label{eq:chi_geom}
    \chi^{0:\mu\mu}_\mathrm{m:geom} = \int_{\mathrm{BZ}} \left(5\cos^2\theta(\mathbf{k}) - 2\right) F[E_{1}(\mathbf{k}), E_{2}(\mathbf{k})]g_{\mu\mu},
\end{align}
where $v_{n,\mu}$ is the Fermi velocity and $F[E_{1}, E_{2}] = f'[E_{1}] + f'[E_{2}] - 2\left(f[E_{2}] - f[E_{1}] \right)/\left(E_2 - E_1 \right)$. The band index of the quantum metric $g_{\mu\mu}$ is omitted, because it is the same for both bands at any momentum. 

It is insightful to compare above equations to previous results for ferromagnetic fluctuations, that is the expressions for the GES from Ref.\ \cite{taisei}. In contrast to $\chi^{0:\mu\mu}_\mathrm{e}$, the effect of the local geometry on $\chi^{0:\mu\mu}_\mathrm{m}$ is not solely due to the explicit dependence on the quantum metric $g$, but also enters through the Bloch angle $\theta$. In fact, we find that $\theta$ is related to the interband Berry connection or dipole coupling\cite{Uzan-Narovlansky2024}, and hence the quantum metric itself can be written as
\begin{align}
    g_{\mu\mu}=\frac{1}{2}\theta^\mu(\mathbf{k})\theta^\mu(\mathbf{k})
\end{align}
where we denote the derivative of $\theta$ w.r.t. $k_\mu$ as $\theta^\mu$. In the vicinity of a (near) degeneracy, the altermagnetic spin-splitting term $t_z$ is absent, and $\cos \theta \rightarrow 1$ and $\sin \theta \rightarrow 0$. There, we recover the expression for the GES: $\chi^{0:\mu\mu}_\mathrm{m}\rightarrow \chi^{0:\mu\mu}_\mathrm{e}$. Away from an intersection (finite spin-splitting term $t_z$), the altermagnetic ($\tau_{\mathrm{M}}=\mathbbm{1}_{2\times 2}$) and ferromagnetic fluctuations ($\tau_{\mathrm{M}}=\tau_z$) are discriminated by their (in-)dependence on the local geometry through $\theta$ across the BZ. For $t_z\rightarrow0$ (going from an altermagnet to a conventional anti-ferromagnet) only $\chi^{0:\mu\mu}_\mathrm{m:mass}$ survives and since $\cos^2(\theta)\rightarrow1$ the GMS is completely non-geometric. 

The sign, but in particular the magnitude of the individual terms in $\chi^{0:\mu\mu}_\mathrm{m}$ at a given momentum vary strongly with the details of the dispersion relation, chemical potential and temperature, because -- just like the quantum metric -- the Fermi-Dirac distribution and its derivatives are ``spiky". Ref.\ \cite{taisei} showed that $\chi^{0:\mu\mu}_\mathrm{e:geom}$ can be negative (favouring FM instabilities) and dominating at low temperatures if the chemical potential is placed within an avoided crossing. We do not require such fine-tuning because -- even though $\chi^{0:\mu\mu}_\mathrm{m:mass}$ will typically dominate -- finite quantum metric $g$ implies $\chi^{\mathrm{AM}} > \chi^{\mathrm{FM}}$. Moreover, $\chi^{0:\mu\mu}_\mathrm{m:geom}$ is negative (favouring an AM instability) in the vicinity of (near) degeneracies as long as the considered band-pair is below the Fermi level since the quantum metric is positive semidefinite and $F[E_1,E_2]\leq 0$ if $E_1, E_2 < 0$ for any temperature. 


\begin{figure}
    \centering
    \includegraphics[width=\linewidth]{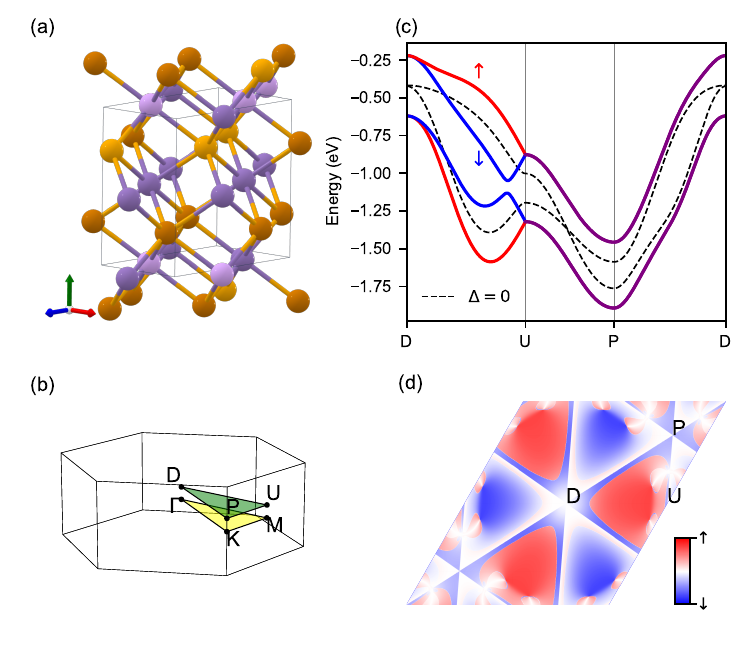}
    \caption{(a) The crystal structure of MnTe taken from Ref.\ \cite{materialsproject}. Planes spanned by a layer of Te atoms (orange) are mirror symmetries that generate the orbit of the Mn atoms (purple). (b) The first BZ of MnTe. The plane spanned by $\Gamma$MK$\Gamma$ (yellow) is a mirror plane of the system. The off-mirror DUPD-plane (green) is shifted by $0.3\pi/c$ along $k_z$. (c) Band structure along D-U-P-D of the normal $\Delta=0$ (dashed) and the altermagnetic state with $\Delta=0.2$. Along D-U, the degeneracy of up- (red) and down-spins (blue) is lifted. Along U-P-D (purple), mirror symmetries preserve the Kramers' degeneracy. (d) The spin expectation value in the DUPD-plane of the second lowest band. The $g$-wave pattern alternates between positive (red) and negative values (blue).}
    \label{fig:figure1}
\end{figure}

\section{Example} The first experimental evidence of altermagnetism  was found in $\alpha$-MnTe\cite{mnte_prl,mnte_nature}. In Fig.\ \ref{fig:figure1}(a) we show its hexagonal NiAs-type structure crystalizing in space group symmetry $P6_3/mmc$ (\# 194) with Mn$^{2+}$ on Wyckoff position $2a$ and Te$^{2-}$ on $2c$\cite{kriegner_mnte,materialsproject}. It is an insulator and shares the space group symmetry with the metallic altermagnet CrSb\cite{crsb}. The site-symmetry (stabilizer) group of Wyckoff position $2a$ is $D_{3d}$ (which contains inversion) and its orbit is generated by a mirror or six-fold screw rotation\cite{bilbao1,bilbao2}. MnTe orders antiferromagnetically below a N\'eel temperature $T_N = 307 K$ with the Mn spins aligning in the $ab$-plane antiferromagnetically between adjacent layers, and in-plane along the (210)-direction (along a Mn-Mn bond)\cite{kriegner_mnte}.

To construct a model of the form Eq.\ \ref{eq:normalstate_hamiltonian}, we perform relativistic ab-initio calculations of N\'eel-ordered MnTe using a generalized-gradient approximation\cite{perdew1996generalized,PBE} as implemented within the VASP package\cite{vasp1,vasp2,vasp3} and then fit a two-site tight-binding model using the topwave package\cite{topwave}. The magnetic sublattice (only the Mn sites) has higher symmetry than the real crystal. In order for the true space group symmetry to become manifest, long-range hoppings are needed. The first Mn-Mn exchange path that discriminates between the altermagnetic, and what would otherwise be an anti-ferromagnetic symmetry are the tenth- and eleventh-nearest neighbors\cite{split_magnon}. The full details of the model can be found in the supplemental material\cite{supp}. The $k_z=0$ mirror enforces a Kramers' degeneracy for zero and $k_z=\pi/c$ out-of-plane momenta, so in Fig.\ \ref{fig:figure1}(b) we show the (non-)magnetic band structure of the effective model along a path of constant $k_z=0.3\pi/c$ as shown in Fig.\ \ref{fig:figure1}(b)). The bands are spin-split along D-U and in Fig.\ \ref{fig:figure1}(d) we plot the spin expectation values of the second lowest band that has the characteristic $g$-wave pattern of N\'eel-ordered MnTe. 

\begin{figure}
    \centering
    \includegraphics[width=\linewidth]{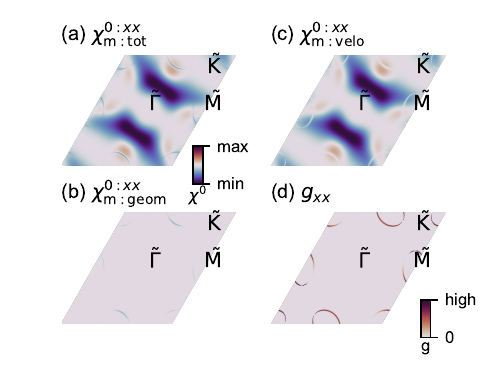}
    \caption{(a) The total generalized magnetic susceptibility $\chi^{0:xx}_\mathrm{m:tot}$ of the normal state model at $\beta=3.7$ close to the $\Gamma$MK$\Gamma$ plane \mbox{(see Fig.\ \ref{fig:figure1}(b))}. (b) The geometric part $\chi^{0:xx}_\mathrm{m:geom}$ of (a). It contributes to the circular lines at the boundary of the BZ. (c) $\chi^{0:xx}_\mathrm{m:velo}$, which makes up the wing-shaped features in the middle of the BZ. (a)-(c) share normalization. (d) The quantum metric $g_{xx}$. Its distribution across the plane has a similar shape as (b).}
    \label{fig:figure2}
\end{figure}

Next, we compute Eq.\ \ref{eq:chi_mass}-\ref{eq:chi_geom} for the non-magnetic model ($\Delta=0$) and plot them in Fig.\ \ref{fig:figure2}(a)-(c) for a $k_z$-cut close to the $\Gamma$MK$\Gamma$-plane shown in Fig.\ \ref{fig:figure1}(b) on a $1000^2$-grid. MnTe is an insulator, and with both bands below the Fermi level, we find $\chi^{0:xx}_\mathrm{m:tot}$ to be overwhelmingly negative. We show the quantum metric $g_{xx}$ in Fig. \ref{fig:figure2}(d) and note that $\chi^{0:xx}_\mathrm{m:geom}$ is large where $g_{xx}$ is large. As predicted, these are the regions where $|E_1 - E_2|$ is smallest. There, $\chi^{0:xx}_\mathrm{m:mass}$ peaks too, because $\cos \theta \rightarrow 1$ and 
\begin{align}
    \chi^{0:\mu\mu}_\mathrm{m:mass} \rightarrow -\frac{1}{6}\int_{\mathrm{BZ}}\sum_{n=1}^{2} f^{(2)}[E_n(\mathbf{k})]\partial_{k_\mu}\partial_{k_\mu}E_n(\mathbf{k})
\end{align}
through integration by parts. $\partial_{k_\mu}\partial_{k_\mu}E_n$ is the inverse effective mass, which is large when a crossing is avoided. 

For temperatures just above $T_{\mathrm{N}}$, the numerical values of $\chi^{0:xx}_\mathrm{m:mass}$ are an order-of-magnitude larger than $\chi^{0:xx}_\mathrm{m:geom}$ (integrated over the BZ). To make the individual contributions of the GMS more easily discernable, the results in Fig.\ \ref{fig:figure2}(a)-(c) were computed at one order-of-magnitude above the N\'eel-temperature. We find that the conclusions of our analysis remain unaffected by these specifics, and remark that the two so-far experimentally verified altermagnets MnTe\cite{mnte_prl,mnte_nature} and CrSb\cite{crsb} have comparatively high $T_{\mathrm{N}}$\cite{am_candidates}. 

\section{Discussion} In this work, we have related the occurrence of altermagnetic spin-splitting and the quantum metric. We have derived an expression for the GMS that elucidates the role of quantum geometry in the formation of altermagnetic instabilities, and have verified our predictions by constructing an effective model for MnTe. Our expression can easily be (numerically) evaluated for other altermagnetic models, and a comparative analysis for other (proposed) altermagnets like RuO$_2$\cite{laura_ruo2}, CrSb\cite{crsb}, FeSb$_2$\cite{fesb2}, etc.\cite{am_candidates} is left for future research. Several theoretical avenues to expand on our work present themselves, including an analysis of the GMS for a more general multi-band model. The extension of the equations found in the supplemental material\cite{supp} are straight-forward in principle, however the identification of the resulting terms with known (geometric) quantities and overall interpretability is non-trivial. Applied to altermagnetic models, we found some of the higher-order terms that arise in the multi-band GMS to cancel. In general, cancellation over the BZ depends on the transformation properties of the different factors under inversion (AM band structures always look inversion symmetric\cite{am_review2}) and the other point group symmetries. Altermagnets have even parity, so any quantity that is odd under a symmmetry that exchanges the sites vanishes over the BZ. The particular symmetries of an altermagnet were shown to constrain transport\cite{cano_am_transport_nonlinear} and they can restrict $\chi^{0:\mu\nu}_\mathrm{m}$ too. The exact implications of these constraints depend on the specific spin-point group of a given altermagnet, and have to be evaluated case-by-case. 

Secondly, we wish to investigate the effect of different interactions on $\chi^{\mathrm{AM}}$ in the future, specifically in models where FM and N\'eel-type order compete. Within a simple random phase approximation (RPA) treatment of on-site interactions, the prominent peak of the non-interacting susceptibility gets enhanced most\cite{taisei,roig2024minimal}, however beyond RPA longer-range interactions might compete with the quantum metric. For example, it was recently proposed that strong interactions can reduce the altermagnetic spin gap in itinerant systems\cite{giuli2024altermagnetisminteractiondrivenitinerantmagnetism}.

We believe that our work inspires fruitful research in that direction and hope that it motivates the experimental community to consider altermagnets as platforms to measure signatures of quantum geometry through non-linear optical response and transport experiments\cite{measure1,measure2,measure3,measure4,measure5,measure6}. 

\begin{acknowledgements}
    We thank K. Alpin, L. Debbeler, T. Kitamura, M. Franz and A. Schnyder for helpful discussions and I. Elfimov for support with the ab-initio calculations. NH acknowledges financial support from the Max Planck Institute for Solid State Research in Stuttgart and the Quantum Electronic Science and Technology program from the Quantum Matter Institute at the University of British Columbia. \\
    \textit{Note added.} We thank Prof. Min-Fong Yang who brought an error in the derivation of the GMS to our attention. An incorrect Matsubara sum and a typographical lead to wrong weighting factors in Eq.~\ref{eq:chi_velo} and Eq.~\ref{eq:chi_geom}. These
    corrections do not affect any of the discussion or conclusions of the paper. We have submited an erratum to the original publication and updated Eq.~\ref{eq:chi_velo} and Eq.~\ref{eq:chi_geom} and the supplemental material.
\end{acknowledgements}

\nocite{matsubara}

\bibliography{apssamp}

\onecolumngrid
\subsection{Response Functions of Two-Band Models}
\noindent 
For a generic two-band model, we define the particle density operator for (spinless) fermions at imaginary time $\tau$ as 
\begin{align}
    n(\tau, \mathbf{q}) = \frac{1}{N}\sum_\mathbf{k}\mathbf{c}^\dagger(\tau,\mathbf{k})\mathbbm{1}_{2\times 2}\mathbf{c}(\tau,\mathbf{k} + \mathbf{q})
\end{align}
with $\mathbf{c}(\tau,\mathbf{k}) = (c_1(\tau,\mathbf{k}), c_2(\tau,\mathbf{k}))^\top$. The subscript of the fermionic annihilation operator $c_\alpha(\tau,\mathbf{k})$ corresponds to the two sites in the unit cell $\alpha = 1, 2$. The susceptibility is given by 
\begin{align}
    \chi(i\Omega_n, \mathbf{q}) = \int_0^\beta d\tau e^{i\Omega_n \tau} \langle T_\tau  n(\tau, \mathbf{q}) n(0, \mathbf{-q})\rangle
\end{align}
with Matsubara frequency $i\Omega_n$, the inverse temperature $\beta$ and the time-ordering operator $T_\tau$. We define a matrix $C$ to collect all time-ordered four operator terms
\begin{align}
    \left[C(\tau, \mathbf{q})\right]_{\alpha\alpha'} = \sum_{\mathbf{k},\mathbf{k}'}\langle T_\tau  c_\alpha^\dagger(\tau, \mathbf{k})c_\alpha(\tau, \mathbf{k}+\mathbf{q}) c_{\alpha'}^\dagger(0, \mathbf{k}')c_{\alpha'}(0, \mathbf{k}'-\mathbf{q}) \rangle,
\end{align}
and rewrite the susceptibility as a sum over all its entries
\begin{align}
    \chi(i\Omega_n, \mathbf{q}) = \frac{1}{N^2} \int_0^\beta d\tau e^{i\Omega_n \tau} \sum_{\alpha,\alpha'} \left[C(\tau, \mathbf{q})\right]_{\alpha\alpha'}.
\end{align}
Next, we introduce a modified density operator 
\begin{align}
    \tilde{n}(\tau, \mathbf{q}) = \frac{1}{N}\sum_\mathbf{k}\mathbf{c}^\dagger(\tau,\mathbf{k})\tau_z\mathbf{c}(\tau,\mathbf{k} + \mathbf{q})
\end{align}
that we use to define the altermagnetic susceptibility as given in Eq. (13) in \cite{roig2024minimal}
\begin{align}
    \chi^{\mathrm{AM}}(i\Omega_n, \mathbf{q}) &= \int_0^\beta d\tau e^{i\Omega_n \tau} \langle T_\tau  \tilde{n}(\tau, \mathbf{q}) \tilde{n}(0, \mathbf{-q})\rangle.
\end{align}
For the altermagnetic susceptibility, all $\alpha \neq \alpha'$ terms acquire a minus sign, so it can be written as
\begin{align}
\chi^{\mathrm{AM}}(i\Omega_n, \mathbf{q}) = \frac{1}{N^2} \int_0^\beta d\tau e^{i\Omega_n \tau} \sum_{\alpha,\alpha'} \left[\tau_z C(\tau, \mathbf{q})\tau_z\right]_{\alpha\alpha'}.
\end{align}
We want to express the susceptibility as a product of Greens function. To this end, we first rearrange the annihilation operators in the matrix elements of $C$, and then apply Wick's theorem:
\begin{align}
     &\sum_{\mathbf{k},\mathbf{k}'}\langle T_\tau  c_\alpha^\dagger(\tau, \mathbf{k})c_\alpha(\tau, \mathbf{k}+\mathbf{q}) c_{\alpha'}^\dagger(0, \mathbf{k}')c_{\alpha'}(0, \mathbf{k}'-\mathbf{q}) \rangle\nonumber \\
     =-&\sum_{\mathbf{k},\mathbf{k}'}\langle T_\tau c_\alpha(\tau, \mathbf{k}+\mathbf{q}) c_{\alpha'}(0, \mathbf{k}'-\mathbf{q})  c_\alpha^\dagger(\tau, \mathbf{k})c_{\alpha'}^\dagger(0, \mathbf{k}')\rangle\nonumber \\
     =-&\sum_{\mathbf{k},\mathbf{k}'}{\big[}\langle T_\tau c_\alpha(\tau, \mathbf{k}+\mathbf{q})c_{\alpha'}^\dagger(0, \mathbf{k}')\rangle \langle T_\tau  c_{\alpha'}(0, \mathbf{k}'-\mathbf{q})  c_\alpha^\dagger(\tau, \mathbf{k})\rangle -\langle T_\tau c_\alpha(\tau, \mathbf{k}+\mathbf{q})c_{\alpha}^\dagger(\tau, \mathbf{k})\rangle \langle T_\tau  c_{\alpha'}(0, \mathbf{k}'-\mathbf{q})  c_{\alpha'}^\dagger(0, \mathbf{k}')\rangle{\big]}\nonumber \\
     =-&\sum_{\mathbf{k},\mathbf{k}'}\delta^{(3)}(\mathbf{k} + \mathbf{q} - \mathbf{k}')\delta^{(3)}(\mathbf{k}' - \mathbf{q} - \mathbf{k}) G^{\alpha \alpha'}(\tau, \mathbf{k}+\mathbf{q})G^{\alpha' \alpha}(-\tau, \mathbf{k})\nonumber \\
     =-N&\sum_{\mathbf{k}} G^{\alpha \alpha'}(\tau, \mathbf{k}+\mathbf{q})G^{\alpha' \alpha}(-\tau, \mathbf{k})
\end{align}
We disregarded the disconnected bubbles that have no frequency dependence (third line), imposed momentum conservation, and used the definition of the imaginary time Greens function
\begin{align}
    G(\tau - \tau', \mathbf{k}-\mathbf{k}') = \langle T_\tau c(\tau, \mathbf{k})c^\dagger(\tau', \mathbf{k}') \rangle.
\end{align}
In this new form, we can write $C$ as
\begin{align}
    C(\tau, \mathbf{q}) = -N\sum_{\mathbf{k}}\begin{pmatrix}
    G^{11}(\tau, \mathbf{k}+\mathbf{q})G^{11}(-\tau, \mathbf{k}) & G^{12}(\tau, \mathbf{k}+\mathbf{q})G^{21}(-\tau, \mathbf{k})\\
    G^{21}(\tau, \mathbf{k}+\mathbf{q})G^{12}(-\tau, \mathbf{k}) & G^{22}(\tau, \mathbf{k}+\mathbf{q})G^{22}(-\tau, \mathbf{k})
    \end{pmatrix}.
\end{align}
It is useful to write the sum over the matrix elements of $C$ as a trace of a matrix product
\begin{align}
     \sum_{\alpha,\alpha'} \left[C(\tau, \mathbf{q})\right]_{\alpha\alpha'}= -N&\sum_{\mathbf{k}}\mathrm{Tr}\left[\mathcal{G}(\tau, \mathbf{k}+\mathbf{q})\mathcal{G}(-\tau, \mathbf{k})\right]\nonumber \\= -N&\sum_{\mathbf{k}}\mathrm{Tr} \left[\begin{pmatrix}
    G^{11}(\tau, \mathbf{k}+\mathbf{q}) & G^{12}(\tau, \mathbf{k}+\mathbf{q})\\
    G^{21}(\tau, \mathbf{k}+\mathbf{q}) & G^{22}(\tau, \mathbf{k}+\mathbf{q})
    \end{pmatrix}\begin{pmatrix}
    G^{11}(-\tau, \mathbf{k}) & G^{12}(-\tau, \mathbf{k})\\
    G^{21}(-\tau, \mathbf{k}) & G^{22}(-\tau, \mathbf{k})
    \end{pmatrix}\right].
\end{align}
The expression for the susceptibility becomes
\begin{align}
    \chi^{\mathrm{(AM)}}(i\Omega_n, \mathbf{q}) = -\frac{1}{N} \int_0^\beta d\tau e^{i\Omega_n \tau} \sum_{\mathbf{k}} \mathrm{Tr}\left[\mathcal{G}(\tau, \mathbf{k}+\mathbf{q})\tau_\mathrm{M} \mathcal{G}(-\tau, \mathbf{k})\tau_\mathrm{M}\right]\label{eq:chi_imag_time}
\end{align}
with $\tau_\mathrm{M}=\mathbbm{1}_{2\times 2}$ or $\tau_\mathrm{M}=\tau_z$ for the susceptibility $\chi$ or the altermagnetic susceptibility $\chi^{\mathrm{AM}}$ respectively. The Fourier transform of the imaginary time Greens function is given by
\begin{align}
    G(\tau, \mathbf{k}) = \frac{1}{\beta}\sum_m e^{-i\omega_m\tau} G(i\omega_m, \mathbf{k}).
\end{align}
Inserting the Fourier transformed Greens function into Eq. \ref{eq:chi_imag_time} above and performing the imaginary time integration we find
\begin{align}\label{eq:greens_function_tau_M}
    \chi^{\mathrm{(AM)}}(i\Omega_n, \mathbf{q}) &= -\frac{1}{N\beta^2} \sum_{m,m'}\int_0^\beta d\tau e^{(i\Omega_n + i\omega_m - i\omega_{m'}) \tau} \sum_{\mathbf{k}} \mathrm{Tr}\left[\mathcal{G}(i\omega_{m'}, \mathbf{k}+\mathbf{q})\tau_\mathrm{M} \mathcal{G}(i\omega_m, \mathbf{k})\tau_\mathrm{M}\right]\nonumber \\
    &= -\frac{1}{N\beta} \sum_{m}\sum_{\mathbf{k}} \mathrm{Tr}\left[\mathcal{G}(i\Omega_n + i\omega_{m}, \mathbf{k}+\mathbf{q})\tau_\mathrm{M} \mathcal{G}(i\omega_m, \mathbf{k})\tau_\mathrm{M}\right].
\end{align}
\subsection{$\mathbf{q}=0$ Instabilities in noninteracting systems}
\noindent The generalized electric susceptibility (GES) is a thermodynamic quanity that is defined as 
\begin{align}
    \chi^{0:\mu\nu}_\mathrm{e} = \lim_{\mathbf{q}\rightarrow 0} \partial_{q_\mu}\partial_{q_\nu}\chi(0, \mathbf{q}).
\end{align}
In analogy to Ref.\ \cite{taisei}, we define the generalized magnetic susceptibility as 
\begin{align}
    \chi^{0:\mu\nu}_\mathrm{m} = \lim_{\mathbf{q}\rightarrow 0} \partial_{q_\mu}\partial_{q_\nu}\chi^{\mathrm{(AM)}}(0, \mathbf{q}).
\end{align}
We note that $\partial_{q_\mu}G(i\omega_m, \mathbf{k}+\mathbf{q}) = \partial_{k_\mu}G(i\omega_m, \mathbf{k}+\mathbf{q})$ and insert our expression for $\chi$ into the equation above
\begin{align}
    \lim_{\mathbf{q}\rightarrow 0} \partial_{q_\mu}\partial_{q_\nu}\chi^{\mathrm{(AM)}}(0, \mathbf{q}) &=  -\lim_{\mathbf{q}\rightarrow 0} \frac{1}{N\beta} \sum_{m}\sum_{\mathbf{k}} \mathrm{Tr}\left[\partial_{q_\mu}\partial_{q_\nu}\mathcal{G}(i\omega_{m}, \mathbf{k}+\mathbf{q})\tau_\mathrm{M} \mathcal{G}(i\omega_m, \mathbf{k})\tau_\mathrm{M}\right]\nonumber \\
    &=  -\lim_{\mathbf{q}\rightarrow 0} \frac{1}{N\beta} \sum_{m}\sum_{\mathbf{k}} \mathrm{Tr}\left[\partial_{k_\mu}\partial_{k_\nu}\mathcal{G}(i\omega_{m}, \mathbf{k}+\mathbf{q})\tau_\mathrm{M} \mathcal{G}(i\omega_m, \mathbf{k})\tau_\mathrm{M}\right]\nonumber \\
    &= -\frac{1}{N\beta} \sum_{m}\sum_{\mathbf{k}} \mathrm{Tr}\left[\partial_{k_\mu}\partial_{k_\nu}\mathcal{G}(i\omega_{m}, \mathbf{k})\tau_\mathrm{M} \mathcal{G}(i\omega_m, \mathbf{k})\tau_\mathrm{M}\right]\nonumber \\
    &= \frac{1}{N\beta} \sum_{m}\sum_{\mathbf{k}} \mathrm{Tr}\left[\partial_{k_\mu}\mathcal{G}(i\omega_{m}, \mathbf{k})\tau_\mathrm{M} \partial_{k_\nu}\mathcal{G}(i\omega_m, \mathbf{k})\tau_\mathrm{M}\right].
\end{align}
In the last step, we used partial integration and made use of the fact the boundary term vanishes, because the sum over $\mathbf{k}$ runs over the whole Brillouin zone. For noninteracting systems 
\begin{align}\label{eq:nonint_greens}\mathcal{G}(i\omega_m,\mathbf{k}) = \left(i\omega_m - \mathcal{H}_0(\mathbf{k})\right)^{-1}.\end{align}
For now, the exact form of $\mathcal{H}_0$ does not matter, except that it is a (spinless) $2\times 2$-matrix with $\mathbf{c}(\tau,\mathbf{k}) = (c_1(\tau,\mathbf{k}), c_2(\tau,\mathbf{k}))^\top$ as a basis. Later, we will identify it with the minimal altermagnetic model in its normal state as given in the main text, and make use of its symmetries. 

Using Eq.\ \ref{eq:nonint_greens}, we now relate the Greens function to the spectrum of the noninteracting Hamiltonian $\mathcal{H}_0$ by inserting two identities and using the cyclic property of the trace:
\begin{align}
    \lim_{\mathbf{q}\rightarrow 0} \partial_{q_\mu}\partial_{q_\nu}\chi^{\mathrm{(AM)}}(0, \mathbf{q}) &= \frac{1}{N\beta} \sum_{m}\sum_{\mathbf{k}} \mathrm{Tr}\left[\partial_{k_\mu}\frac{1}{i\omega_m - \mathcal{H}_0(\mathbf{k})}\tau_\mathrm{M} \partial_{k_\nu}\frac{1}{i\omega_m - \mathcal{H}_0(\mathbf{k})}\tau_\mathrm{M}\right]\nonumber  \\
    &= \frac{1}{N\beta} \sum_{m}\sum_{n,\tilde{n}}\sum_{\mathbf{k}} \langle u_n(\mathbf{k})|\partial_{k_\mu}\frac{1}{i\omega_m - \mathcal{H}_0(\mathbf{k})}\tau_\mathrm{M}|u_{\tilde{n}}(\mathbf{k})\rangle\langle u_{\tilde{n}}(\mathbf{k})| \partial_{k_\nu}\frac{1}{i\omega_m - \mathcal{H}_0(\mathbf{k})}\tau_\mathrm{M}|u_n(\mathbf{k})\rangle.\label{eq:GES_hamiltonian}
\end{align}

The case of $\tau_\mathrm{M}=\mathbbm{1}_{2\times 2}$ can be found in the supplemental material of Ref.\ \cite{taisei}. Here, we are going to focus on the altermagnetic susceptibility with $\tau_\mathrm{M}=\tau_z$. For an altermagnet, $\tau_z$ and $\mathcal{H}_0$ do not commute, so we insert two more identities into Eq.\ \ref{eq:GES_hamiltonian}:
\begin{align}
    \frac{1}{N\beta} \sum_{m}\sum_{n,n'}\sum_{\tilde{n},\tilde{n}'}\sum_{\mathbf{k}} \langle u_n(\mathbf{k})|\partial_{k_\mu}\frac{1}{i\omega_m - \mathcal{H}_0(\mathbf{k})}|u_{n'}(\mathbf{k})\rangle\langle u_{n'}(\mathbf{k})|\tau_z|u_{\tilde{n}}(\mathbf{k})\rangle\langle u_{\tilde{n}}(\mathbf{k})| \partial_{k_\nu}\frac{1}{i\omega_m - \mathcal{H}_0(\mathbf{k})}|u_{\tilde{n}'}(\mathbf{k})\rangle\langle u_{\tilde{n}'}(\mathbf{k})|\tau_z|u_n(\mathbf{k})\rangle \label{eq:GES_more_identities}.
\end{align}
The next step is to replace the matrix $\mathcal{H}_0$ in the denominators to explicitly perform the Matsubara sums. We note that the partial derivative of the Greens function can be written as $\partial_{k_\mu}\mathcal{G}(i\omega_m,\mathbf{k}) = -\mathcal{G}(i\omega_m,\mathbf{k})\partial_{k_\mu}\mathcal{H}_0(\mathbf{k})\mathcal{G}(i\omega_m,\mathbf{k})$, because
\begin{align}
    \partial_{k_\mu}\mathbbm{1} &= \partial_{k_\mu}\left[\left(i\omega_m - \mathcal{H}_0(\mathbf{k})\right)\left(i\omega_m - \mathcal{H}_0(\mathbf{k})\right)^{-1}\right]\nonumber \\
    0 &= \partial_{k_\mu}\left[\left(i\omega_m - \mathcal{H}_0(\mathbf{k})\right)\right]\left(i\omega_m - \mathcal{H}_0(\mathbf{k})\right)^{-1} + \left(i\omega_m - \mathcal{H}_0(\mathbf{k})\right)\partial_{k_\mu}\left[\left(i\omega_m - \mathcal{H}_0(\mathbf{k})\right)^{-1}\right]\nonumber \\
    0 &= \left(i\omega_m - \mathcal{H}_0(\mathbf{k})\right)^{-1}\partial_{k_\mu}\left[\left(i\omega_m - \mathcal{H}_0(\mathbf{k})\right)\right]\left(i\omega_m - \mathcal{H}_0(\mathbf{k})\right)^{-1} + \partial_{k_\mu}\left[\left(i\omega_m - \mathcal{H}_0(\mathbf{k})\right)^{-1}\right] \nonumber \\
    0 &= -\mathcal{G}(i\omega_m,\mathbf{k})\partial_{k_\mu}\mathcal{H}_0(\mathbf{k})\mathcal{G}(i\omega_m,\mathbf{k}) + \partial_{k_\mu}\mathcal{G}(i\omega_m,\mathbf{k}).
\end{align}
Using, $\left(i\omega_m - \mathcal{H}_0(\mathbf{k})\right)^{-1}|u_n(\mathbf{k})\rangle = \left(i\omega_m - E_n(\mathbf{k})\right)^{-1}|u_n(\mathbf{k})\rangle$, we rewrite the matrix elements in Eq.\ \ref{eq:GES_more_identities} as
\begin{align}
    \langle u_n(\mathbf{k})|\partial_{k_\mu}\frac{1}{i\omega_m - \mathcal{H}_0(\mathbf{k})}|u_{\tilde{n}}(\mathbf{k})\rangle &= \langle u_n(\mathbf{k})|\frac{1}{i\omega_m - \mathcal{H}_0(\mathbf{k})}\partial_{k_\mu}\mathcal{H}_0(\mathbf{k})\frac{1}{i\omega_m - \mathcal{H}_0(\mathbf{k})}|u_{\tilde{n}}(\mathbf{k})\rangle \nonumber\\\label{eq:deriv_G_matel}
    &= \frac{\langle u_n(\mathbf{k})|\partial_{k_\mu}\mathcal{H}_0(\mathbf{k})|u_{\tilde{n}}(\mathbf{k})\rangle}{\left(i\omega_m - E_n(\mathbf{k})\right)\left(i\omega_m - E_{\tilde{n}}(\mathbf{k})\right)}.
\end{align}
Inserting this into Eq. \ref{eq:GES_more_identities} yields
\begin{align}\frac{1}{N\beta} \sum_{m}\sum_{n,n'}\sum_{\tilde{n},\tilde{n}'}\sum_{\mathbf{k}} \frac{\langle u_n(\mathbf{k})|\partial_{k_\mu}\mathcal{H}_0(\mathbf{k})|u_{n'}(\mathbf{k})\rangle}{\left(i\omega_m - E_n(\mathbf{k})\right)\left(i\omega_m - E_{n'}(\mathbf{k})\right)}\langle u_{n'}(\mathbf{k})|\tau_z|u_{\tilde{n}}(\mathbf{k})\rangle\frac{\langle u_{\tilde{n}}(\mathbf{k})| \partial_{k_\nu}\mathcal{H}_0(\mathbf{k})|u_{\tilde{n}'}(\mathbf{k})\rangle}{\left(i\omega_m - E_{\tilde{n}}(\mathbf{k})\right)\left(i\omega_m - E_{\tilde{n}'}(\mathbf{k})\right)}\langle u_{\tilde{n}'}(\mathbf{k})|\tau_z|u_n(\mathbf{k})\rangle.
\end{align}
For bookkeeping, we now organize the quadruple sum over the band indices into separate sums (if there is no equal sign, it means the indices are strictly unequal): \\\\(i)\ \underline{$n=n'=\tilde{n}=\tilde{n}'$}:
\begin{align}
    \frac{1}{N\beta} \sum_{m}\sum_{n}\sum_{\mathbf{k}} \frac{\langle u_n(\mathbf{k})|\partial_{k_\mu}\mathcal{H}_0(\mathbf{k})|u_{n}(\mathbf{k})\rangle\langle u_{{n}}(\mathbf{k})| \partial_{k_\nu}\mathcal{H}_0(\mathbf{k})|u_{{n}}(\mathbf{k})\rangle}{\left(i\omega_m - E_n(\mathbf{k})\right)^4}\left|\langle u_{n}(\mathbf{k})|\tau_z|u_{{n}}(\mathbf{k})\rangle\right|^2
\end{align}
(ii)\ \underline{$n={n}'$, $\tilde{n}=\tilde{n}'$}:
\begin{align}
    \frac{1}{N\beta} \sum_{m}\sum_{n\neq\tilde{n}}\sum_{\mathbf{k}} \frac{\langle u_n(\mathbf{k})|\partial_{k_\mu}\mathcal{H}_0(\mathbf{k})|u_{n}(\mathbf{k})\rangle}{\left(i\omega_m - E_n(\mathbf{k})\right)^2}\frac{\langle u_{\tilde{n}}(\mathbf{k})|\partial_{k_\nu}\mathcal{H}_0(\mathbf{k})|u_{\tilde{n}}(\mathbf{k})\rangle}{\left(i\omega_m - E_{\tilde{n}}(\mathbf{k})\right)^2}\left|\langle u_{n}(\mathbf{k})|\tau_z|u_{\tilde{n}}(\mathbf{k})\rangle\right|^2
\end{align}
(iii)\ \underline{$n=\tilde{n}'$, $n'=\tilde{n}$}:
\begin{align}
    \frac{1}{N\beta} \sum_{m}\sum_{n\neq\tilde{n}}\sum_{\mathbf{k}} \frac{\langle u_n(\mathbf{k})|\partial_{k_\mu}\mathcal{H}_0(\mathbf{k})|u_{\tilde{n}}(\mathbf{k})\rangle}{\left(i\omega_m - E_n(\mathbf{k})\right)^2}\frac{\langle u_{\tilde{n}}(\mathbf{k})|\partial_{k_\nu}\mathcal{H}_0(\mathbf{k})|u_{{n}}(\mathbf{k})\rangle}{\left(i\omega_m - E_{\tilde{n}}(\mathbf{k})\right)^2}\langle u_{n}(\mathbf{k})|\tau_z|u_{{n}}(\mathbf{k})\rangle\langle u_{\tilde{n}}(\mathbf{k})|\tau_z|u_{\tilde{n}}(\mathbf{k})\rangle
\end{align}
(iv)\ \underline{$n=n'=\tilde{n}$}:
\begin{align}
    \frac{1}{N\beta} \sum_{m}\sum_{n\neq\tilde{n}'}\sum_{\mathbf{k}} \frac{\langle u_n(\mathbf{k})|\partial_{k_\mu}\mathcal{H}_0(\mathbf{k})|u_{{n}}(\mathbf{k})\rangle}{\left(i\omega_m - E_n(\mathbf{k})\right)^3}\frac{\langle u_{{n}}(\mathbf{k})|\partial_{k_\nu}\mathcal{H}_0(\mathbf{k})|u_{\tilde{n}'}(\mathbf{k})\rangle}{\left(i\omega_m - E_{\tilde{n}'}(\mathbf{k})\right)}\langle u_{n}(\mathbf{k})|\tau_z|u_{{n}}(\mathbf{k})\rangle\langle u_{\tilde{n}'}(\mathbf{k})|\tau_z|u_{n}(\mathbf{k})\rangle
\end{align}
(v)\ \underline{$n=n'=\tilde{n}'$}:
\begin{align}
    \frac{1}{N\beta} \sum_{m}\sum_{n\neq\tilde{n}}\sum_{\mathbf{k}} \frac{\langle u_n(\mathbf{k})|\partial_{k_\mu}\mathcal{H}_0(\mathbf{k})|u_{{n}}(\mathbf{k})\rangle}{\left(i\omega_m - E_n(\mathbf{k})\right)^3}\frac{\langle u_{\tilde{n}}(\mathbf{k})|\partial_{k_\nu}\mathcal{H}_0(\mathbf{k})|u_{{n}}(\mathbf{k})\rangle}{\left(i\omega_m - E_{\tilde{n}}(\mathbf{k})\right)}\langle u_{n}(\mathbf{k})|\tau_z|u_{\tilde{n}}(\mathbf{k})\rangle\langle u_{n}(\mathbf{k})|\tau_z|u_{n}(\mathbf{k})\rangle
\end{align}
(vi)\ \underline{$n=\tilde{n}=\tilde{n}'$}:
\begin{align}
    \frac{1}{N\beta} \sum_{m}\sum_{n\neq n'}\sum_{\mathbf{k}} \frac{\langle u_n(\mathbf{k})|\partial_{k_\mu}\mathcal{H}_0(\mathbf{k})|u_{{n'}}(\mathbf{k})\rangle}{\left(i\omega_m - E_{n'}(\mathbf{k})\right)}\frac{\langle u_{n}(\mathbf{k})|\partial_{k_\nu}\mathcal{H}_0(\mathbf{k})|u_{{n}}(\mathbf{k})\rangle}{\left(i\omega_m - E_{{n}}(\mathbf{k})\right)^3}\langle u_{n'}(\mathbf{k})|\tau_z|u_{{n}}(\mathbf{k})\rangle\langle u_{n}(\mathbf{k})|\tau_z|u_{n}(\mathbf{k})\rangle
\end{align}
(vii)\ \underline{$n'=\tilde{n}=\tilde{n}'$}:
\begin{align}
    \frac{1}{N\beta} \sum_{m}\sum_{n\neq n'}\sum_{\mathbf{k}} \frac{\langle u_n(\mathbf{k})|\partial_{k_\mu}\mathcal{H}_0(\mathbf{k})|u_{{n'}}(\mathbf{k})\rangle}{\left(i\omega_m - E_{n}(\mathbf{k})\right)}\frac{\langle u_{n'}(\mathbf{k})|\partial_{k_\nu}\mathcal{H}_0(\mathbf{k})|u_{{n'}}(\mathbf{k})\rangle}{\left(i\omega_m - E_{{n'}}(\mathbf{k})\right)^3}\langle u_{n'}(\mathbf{k})|\tau_z|u_{{n'}}(\mathbf{k})\rangle\langle u_{n'}(\mathbf{k})|\tau_z|u_{n}(\mathbf{k})\rangle
\end{align}
For each of these sums, we now perform the Matsubara summation, and apply the Hellmann-Feynman theorem
\begin{align}\label{eq:Hellmann-feynmann}
    \langle u_n(\mathbf{k})|\partial_{k_\mu}\mathcal{H}_0(\mathbf{k})|u_{n'}(\mathbf{k})\rangle = \delta_{nn'}\partial_{k_\mu}E_{n}(\mathbf{k}) + (E_n(\mathbf{k}) - E_{n'}(\mathbf{k}))\langle \partial_{k_\mu} u_n(\mathbf{k})|u_{n'}(\mathbf{k})\rangle.
\end{align}

Starting with (i), we find
\begin{align}
    \chi_{\mu\nu}^{\mathrm{(i)}} = \sum_{n}\int\frac{d\mathbf{k}}{(2\pi)^d}\partial_{k_\mu}E_n(\mathbf{k})\partial_{k_\nu}E_n(\mathbf{k}) \frac{f^{(3)}(E_n(\mathbf{k}))}{6}\left|\langle u_{n}(\mathbf{k})|\tau_z|u_{{n}}(\mathbf{k})\rangle\right|^2,
\end{align}
where we substituted the sum over momenta by an integral over the BZ and $f$ as the Fermi-Dirac distribution. This is the same expression as in the supplemental material of Ref.\ \cite{taisei}, but weighted by the diagonal matrix elements of $\tau_z$.

Moving on, (ii) and (iii) share the same Matsubara sum that evaluates to 
\begin{align}
    \frac{1}{\beta}\sum_m \frac{1}{\left(i\omega_m - E_n(\mathbf{k})\right)^2\left(i\omega_m - E_{\tilde{n}}(\mathbf{k})\right)^2} &= \left[ f'(E_n(\mathbf{k})) + f'(E_{\tilde{n}}(\mathbf{k})) + 2\frac{f(E_{\tilde{n}}(\mathbf{k})) - f(E_n(\mathbf{k}))}{E_n(\mathbf{k}) - E_{\tilde{n}}(\mathbf{k})}\right]\frac{1}{\left( E_n(\mathbf{k}) - E_{\tilde{n}}(\mathbf{k})\right)^2} 
\end{align}
Using this, (ii) becomes 
\begin{align}
    \chi_{\mu\nu}^{\mathrm{(ii)}} &= \frac{1}{N} \sum_{\mathbf{k}} \sum_{n\neq\tilde{n}}\partial_{k_\mu}E_n(\mathbf{k})\partial_{k_\nu}E_{\tilde{n}}(\mathbf{k})\bigg(f'(E_n(\mathbf{k})) + f'(E_{\tilde{n}}(\mathbf{k})) +
    2\frac{f(E_{\tilde{n}}(\mathbf{k})) - f(E_{n}(\mathbf{k}))}{E_n(\mathbf{k}) - E_{\tilde{n}}(\mathbf{k})}\bigg)\frac{\left|\langle u_{n}(\mathbf{k})|\tau_z|u_{\tilde{n}}(\mathbf{k})\rangle\right|^2}{\left( E_n(\mathbf{k}) - E_{\tilde{n}}(\mathbf{k})\right)^2}
\end{align}
with the Fermi velocities $v_{n,\mu} = \partial_{k_\mu}E_n(\mathbf{k})$ of a given band $n$ along a direction $\mu$. Next, we want to rewrite $\chi_{\mu\nu}^{(ii)}$ as a sum only over the upper triangular part of band space -- adding the $\tilde{n}<n$ terms explicitly. Because the Fermi functions and matrix elements are invariant under exchanging $n$ and $\tilde{n}$, we can simply factor it out. In this case exchanging the band indices is the same as exchanging $\mu$ and $\nu$, so we find
\begin{align}
    \chi_{\mu\nu}^{\mathrm{(ii)}} &= \int\frac{d\mathbf{k}}{(2\pi)^d} \sum_{n}\sum_{\tilde{n}(>n)}v_{n,\mu}(\mathbf{k})v_{\tilde{n},\nu}(\mathbf{k}) \bigg(\frac{f'(E_n(\mathbf{k})) + f'(E_{\tilde{n}}(\mathbf{k})) }{{\left( E_n(\mathbf{k}) - E_{\tilde{n}}(\mathbf{k})\right)^2}}+
    2\frac{f(E_{\tilde{n}}(\mathbf{k})) - f(E_{n}(\mathbf{k}))}{(E_n(\mathbf{k}) - E_{\tilde{n}}(\mathbf{k}))^3}\bigg)\left|\langle u_{n}(\mathbf{k})|\tau_z|u_{\tilde{n}}(\mathbf{k})\rangle\right|^2 + \mu \leftrightarrow \nu ,
\end{align}
where we replaced the summation over momenta by an integral over the BZ. This is going to be the common strategy for all terms. Evaluate the Matsubara sum (and bring it into a ``nice" form), apply the Hellmann-Feynman theorem, and finally analyze the effect of exchanging $n$ and $\tilde{n}$ to rewrite the sum. 

Next, we repeat the procedure for (iii). We take a closer look at the Hellmann-Feynman theorem applied to the product of matrix elements in (iii). Because $\partial_{k_\mu}\mathcal{H}_0$ is hermitian, it follows that
\begin{align}\label{eq:partial_H_hermitian}
    \frac{\langle u_n(\mathbf{k})|\partial_{k_\mu}\mathcal{H}_0(\mathbf{k})|u_{\tilde{n}}(\mathbf{k})\rangle}{\left(i\omega_m - E_n(\mathbf{k})\right)\left(i\omega_m - E_{\tilde{n}}(\mathbf{k})\right)} = \frac{\langle u_{\tilde{n}}(\mathbf{k})|\partial_{k_\mu}\mathcal{H}_0(\mathbf{k})|u_{n}(\mathbf{k})\rangle^*}{\left(i\omega_m - E_n(\mathbf{k})\right)\left(i\omega_m - E_{\tilde{n}}(\mathbf{k})\right)},
\end{align}
and therefore
\begin{align}
    \frac{\langle u_n(\mathbf{k})|\partial_{k_\mu}\mathcal{H}_0(\mathbf{k})|u_{\tilde{n}}(\mathbf{k})\rangle}{E_n(\mathbf{k}) - E_{\tilde{n}}(\mathbf{k})}\frac{\langle u_{\tilde{n}}(\mathbf{k})|\partial_{k_\nu}\mathcal{H}_0(\mathbf{k})|u_{n}(\mathbf{k})\rangle}{E_n(\mathbf{k}) - E_{\tilde{n}}(\mathbf{k})} &= \frac{\langle u_n(\mathbf{k})|\partial_{k_\mu}\mathcal{H}_0(\mathbf{k})|u_{\tilde{n}}(\mathbf{k})\rangle}{E_n(\mathbf{k}) - E_{\tilde{n}}(\mathbf{k})}\frac{\langle u_{n}(\mathbf{k})|\partial_{k_\nu}\mathcal{H}_0(\mathbf{k})|u_{\tilde{n}}(\mathbf{k})\rangle^*}{E_n(\mathbf{k}) - E_{\tilde{n}}(\mathbf{k})} \nonumber \\
    &=(E_n(\mathbf{k}) - E_{\tilde{n}}(\mathbf{k}))^2\frac{\langle \partial_{k_\mu} u_n(\mathbf{k})|u_{\tilde{n}}(\mathbf{k})\rangle}{E_n(\mathbf{k}) - E_{\tilde{n}}(\mathbf{k})}\frac{\langle \partial_{k_\nu} u_n(\mathbf{k})|u_{\tilde{n}}(\mathbf{k})\rangle^*}{E_n(\mathbf{k}) - E_{\tilde{n}}(\mathbf{k})} \nonumber \\ &= \langle \partial_{k_\mu} u_n(\mathbf{k})|u_{\tilde{n}}(\mathbf{k})\rangle\langle  u_{\tilde{n}}(\mathbf{k})|\partial_{k_\nu}u_{n}(\mathbf{k})\rangle.
\end{align}
Inserting this into (iii), we obtain
\begin{align}
    \chi_{\mu\nu}^{\mathrm{(iii)}} =\frac{1}{N} \sum_{\mathbf{k}}\sum_{n\neq \tilde{n}}\langle \partial_{k_\mu} u_n(\mathbf{k})|u_{\tilde{n}}(\mathbf{k})\rangle\langle  u_{\tilde{n}}(\mathbf{k})|\partial_{k_\nu}u_{n}(\mathbf{k})\rangle \bigg({f'(E_n(\mathbf{k})) + f'(E_{\tilde{n}}(\mathbf{k})) }+ 2\frac{f(E_{\tilde{n}}(\mathbf{k})) - f(E_{n}(\mathbf{k}))}{E_n(\mathbf{k}) - E_{\tilde{n}}(\mathbf{k})}\bigg) \nonumber \\ \times\langle u_{n}(\mathbf{k})|\tau_z|u_{{n}}(\mathbf{k})\rangle\langle u_{\tilde{n}}(\mathbf{k})|\tau_z|u_{\tilde{n}}(\mathbf{k})\rangle.
\end{align}
Again, the Fermi functions and $\tau_z$ matrix elements are invariant under exchange of band indices, and the factors that look like the (interband) Berry connection transform as 
\begin{align}\label{eq:connection_minus}
\langle \partial_{k_\nu}u_{
\tilde{n}
}(\mathbf{k})|u_{n}(\mathbf{k})\rangle = -\langle u_{\tilde{n}}(\mathbf{k})|\partial_{k_\nu}u_{n}(\mathbf{k})\rangle.
\end{align} 
However, they appear in products of two, so the overall sign cancels and we rewrite the sum as 
\begin{align}
    \chi_{\mu\nu}^{\mathrm{(iii)}} =\frac{1}{N} \sum_{\mathbf{k}}\sum_{n} \sum_{\tilde{n}(>n)}&\bigg(\langle \partial_{k_\mu} u_n(\mathbf{k})|u_{\tilde{n}}(\mathbf{k})\rangle\langle  u_{\tilde{n}}(\mathbf{k})|\partial_{k_\nu}u_{n}(\mathbf{k})\rangle + \langle \partial_{k_\mu} u_{\tilde{n}}(\mathbf{k})|u_{{n}}(\mathbf{k})\rangle\langle  u_{{n}}(\mathbf{k})|\partial_{k_\nu}u_{\tilde{n}}(\mathbf{k})\rangle\bigg) \nonumber \\ \times &\bigg({f'(E_n(\mathbf{k})) + f'(E_{\tilde{n}}(\mathbf{k})) }+ 2\frac{f(E_{\tilde{n}}(\mathbf{k})) - f(E_{n}(\mathbf{k}))}{E_n(\mathbf{k}) - E_{\tilde{n}}(\mathbf{k})}\bigg) \langle u_{n}(\mathbf{k})|\tau_z|u_{{n}}(\mathbf{k})\rangle\langle u_{\tilde{n}}(\mathbf{k})|\tau_z|u_{\tilde{n}}(\mathbf{k})\rangle \nonumber \\
    = \int\frac{d\mathbf{k}}{(2\pi)^d}\sum_{n} \sum_{\tilde{n}(>n)}&\langle \partial_{k_\mu} u_n(\mathbf{k})|u_{\tilde{n}}(\mathbf{k})\rangle\langle  u_{\tilde{n}}(\mathbf{k})|\partial_{k_\nu}u_{n}(\mathbf{k})\rangle \langle u_{n}(\mathbf{k})|\tau_z|u_{{n}}(\mathbf{k})\rangle\langle u_{\tilde{n}}(\mathbf{k})|\tau_z|u_{\tilde{n}}(\mathbf{k})\rangle  \nonumber \\ \times &\bigg({f'(E_n(\mathbf{k})) + f'(E_{\tilde{n}}(\mathbf{k})) }+ 2\frac{f(E_{\tilde{n}}(\mathbf{k})) - f(E_{n}(\mathbf{k}))}{E_n(\mathbf{k}) - E_{\tilde{n}}(\mathbf{k})}\bigg)  + \mu \leftrightarrow \nu  
\end{align}
For a general (spinless) multi-band model, $\chi_{\mu\nu}^{\mathrm{(iii)}}$ is proportional to the quantum metric (which we will define and identify later on). We encourage the reader to compare this formula to the results from Ref.\ \cite{taisei}. The authors find that there are two geometric terms -- the quantum metric and the positional shift -- with opposite sign that favor ferro- and anti-ferromagnetic fluctuations respectively. Therein, $\tau_\mathrm{M}=\mathbbm{1}_{2\times 2}$, so only $\chi_{\mu\nu}^{\mathrm{(i)}}$ and $\chi_{\mu\nu}^{\mathrm{(iii)}}$ enter. We will now deal with the terms that are a consequence of having inserted the two extra identities that project $\tau_\mathrm{M}$ onto the band basis.

We sum expressions $(iv)$ to $(vii)$ and write
\begin{align}
    \chi_{\mu\nu}^{(iv)-(vii)} = \frac{1}{N\beta}\sum_m \sum_{n\neq \tilde{n}}\sum_\mathbf{k}\bigg[&\frac{\langle u_n(\mathbf{k}) |\partial_{k_\mu}\mathcal{H}_0(\mathbf{k})|u_n(\mathbf{k}) \rangle \langle u_n(\mathbf{k})|\tau_z |u_n(\mathbf{k})  \rangle}{\left(i\omega_m - E_n(\mathbf{k}) \right)^3 \left( i\omega_m - E_{\tilde{n}}(\mathbf{k})\right)}\nonumber \\ \times \bigg( &\langle u_n(\mathbf{k}) |\partial_{k_\nu}\mathcal{H}_0(\mathbf{k})|u_{\tilde{n}}(\mathbf{k}) \rangle \langle u_{\tilde{n}}(\mathbf{k})|\tau_z |u_n(\mathbf{k})  \rangle + \langle u_{\tilde{n}}(\mathbf{k}) |\partial_{k_\nu}\mathcal{H}_0(\mathbf{k})|u_{{n}}(\mathbf{k}) \rangle \langle u_{{n}}(\mathbf{k})|\tau_z |u_{\tilde{n}}(\mathbf{k})  \rangle \bigg) + \mu \leftrightarrow \nu \bigg].
\end{align}
The Matsubara sum evaluates as follows:
\begin{align}
    \frac{1}{\beta}\sum_{m}\frac{1}{\left(i\omega_m - E_n(\mathbf{k})\right)^3\left(i\omega_m - E_{\tilde{n}}(\mathbf{k})\right)} = \frac{f(E_n(\mathbf{k})) - f(E_{\tilde{n}}(\mathbf{k}))}{\left(E_n(\mathbf{k}) - E_{\tilde{n}}(\mathbf{k})\right)^3} - \frac{f'(E_n(\mathbf{k})) }{\left(E_n(\mathbf{k}) - E_{\tilde{n}}(\mathbf{k})\right)^2} + \frac{f^{(2)}(E_n(\mathbf{k})) }{2\left(E_n(\mathbf{k}) - E_{\tilde{n}}(\mathbf{k})\right)},
\end{align}
and applying Hellmann-Feynman, the expression becomes (using Eq.\ \ref{eq:connection_minus})
\begin{align}\label{eq:chi_fourseven_neqntilde}
    \chi_{\mu\nu}^{(iv)-(vii)} = \frac{1}{N} \sum_{n\neq \tilde{n}}\sum_\mathbf{k}\bigg[ &\langle u_n(\mathbf{k})|\tau_z |u_n(\mathbf{k})  \rangle \partial_{k_\mu}E_n(\mathbf{k})  \bigg( \frac{f(E_n(\mathbf{k})) - f(E_{\tilde{n}}(\mathbf{k}))}{\left(E_n(\mathbf{k}) - E_{\tilde{n}}(\mathbf{k})\right)^2} - \frac{f'(E_n(\mathbf{k})) }{E_n(\mathbf{k}) - E_{\tilde{n}}(\mathbf{k})} + \frac{1}{2}f^{(2)}(E_n(\mathbf{k})) \bigg) \nonumber \\ 
    \times \bigg( &\langle \partial_{k_\nu} u_n(\mathbf{k})|u_{\tilde{n}}(\mathbf{k}) \rangle\langle u_{\tilde{n}}(\mathbf{k})|\tau_z |u_n(\mathbf{k})  \rangle + \langle u_{\tilde{n}}(\mathbf{k}) |\partial_{k_\nu} u_n(\mathbf{k}) \rangle\langle u_{{n}}(\mathbf{k})|\tau_z |u_{\tilde{n}}(\mathbf{k})  \rangle \bigg) + \mu \leftrightarrow \nu \bigg].
\end{align}
This is the first expression that is not invariant under exchange of $n$ and $\tilde{n}$. The Berry-connection-like terms do not appear pairwise, and in general the diagonal matrix elements of $\tau_z$ will have different values for different $n$. Moreover, the terms that contain the Fermi-Dirac distribution and derivatives thereof, are also all odd under exchange of band indices (except for $f^{(2)}/2$). To bring $\chi_{\mu\nu}^{(iv)-(vii)}$ into a nicer form (in particular the factor containing all the Fermi functions), we need an extra sign change when exchanging $n$ and $\tilde{n}$, but the only term that this sign can come from is $\langle u_n(\mathbf{k})|\tau_z |u_n(\mathbf{k})  \rangle$. For the altermagnetic two-band models (and many more), the diagonal elements differ by a minus sign. For a general multi-band model, one would require that pairs of bands are related by a minus sign, but the matrix elements with other bands than its ``partner" are (close to) zero. This scenario is not too unrealistic. For example, in many crystalline topological phases that are protected by mirror symmetries or any symmetry that partners pairs of bands, one is typically interested in the geometric, or interband properties of these band-pairs. In the next section, we will see that for the two-band altermagnetic model this assumption is exact. 

Assuming that $\langle u_n(\mathbf{k})|\tau_z |u_n(\mathbf{k})  \rangle$ have pairwise opposite sign (in $n$) or are zero otherwise, we rewrite the sum as
\begin{align}\label{eq:chi_fourseven_before_product_rule}
    \chi_{\mu\nu}^{(iv)-(vii)} = \frac{1}{N} \sum_n \sum_{\tilde{n}(>n)}\sum_\mathbf{k} \langle u_{{n}}(\mathbf{k})|\tau_z |u_n(\mathbf{k})  \rangle\bigg( &\langle \partial_{k_\nu} u_n(\mathbf{k})|u_{\tilde{n}}(\mathbf{k}) \rangle\langle u_{\tilde{n}}(\mathbf{k})|\tau_z |u_n(\mathbf{k})  \rangle + \langle u_{\tilde{n}}(\mathbf{k}) |\partial_{k_\nu} u_n(\mathbf{k}) \rangle\langle u_{{n}}(\mathbf{k})|\tau_z |u_{\tilde{n}}(\mathbf{k})  \rangle \bigg) \nonumber \\ 
    \times \bigg[ &\left(f(E_n(\mathbf{k})) - f(E_{\tilde{n}}(\mathbf{k}))\right)\frac{\partial_{k_\mu}E_n(\mathbf{k}) - \partial_{k_\mu}E_{\tilde{n}}(\mathbf{k})}{\left(E_n(\mathbf{k}) - E_{\tilde{n}}(\mathbf{k})\right)^2} + \mu \leftrightarrow \nu \nonumber \\
    - &\frac{1}{E_n(\mathbf{k}) - E_{\tilde{n}}(\mathbf{k})}\left(f'(E_{\tilde{n}}(\mathbf{k}))\partial_{k_\mu}E_n(\mathbf{k}) - f'(E_{\tilde{n}}(\mathbf{k}))\partial_{k_\mu}E_{\tilde{n}}(\mathbf{k})\right)+ \mu \leftrightarrow \nu \nonumber \\
    +&\frac{1}{2}\left( f^{(2)}(E_n(\mathbf{k}))\partial_{k_\mu}E_n(\mathbf{k}) + f^{(2)}(E_{\tilde{n}}(\mathbf{k}))\partial_{k_\mu}E_{\tilde{n}}(\mathbf{k}) \right) + \mu \leftrightarrow \nu \bigg].
\end{align}
Applying the chain rule backwards we find that 
\begin{align}
    \left(f(E_n(\mathbf{k})) - f(E_{\tilde{n}}(\mathbf{k})) \right)\frac{\partial_{k_\mu}E_{n}(\mathbf{k}) - \partial_{k_\mu}E_{\tilde{n}}(\mathbf{k})}{\left(E_n(\mathbf{k}) - E_{\tilde{n}}(\mathbf{k})\right)^2} = -\left(f(E_n(\mathbf{k})) - f(E_{\tilde{n}}(\mathbf{k}))\right) \partial_{k_\mu}\frac{1}{E_n(\mathbf{k}) - E_{\tilde{n}}(\mathbf{k})},
\end{align}
and
\begin{align}
    \frac{1}{E_n(\mathbf{k}) - E_{\tilde{n}}(\mathbf{k})} \left(f'(E_{n}(\mathbf{k}))\partial_{k_\mu}E_{n}(\mathbf{k}) - f'(E_{\tilde{n}}(\mathbf{k}))\partial_{k_\mu}E_{\tilde{n}}(\mathbf{k})\right) = \frac{1}{E_n(\mathbf{k}) - E_{\tilde{n}}(\mathbf{k})} \partial_{k_\mu}\left(f(E_{n}(\mathbf{k})) - f(E_{\tilde{n}}(\mathbf{k}))\right).
\end{align}
Now using the product rule backwards on the sum of these two terms, Eq.\ \ref{eq:chi_fourseven_before_product_rule} becomes
\begin{align}\label{eq:sin2theta_afterproductback}
    \chi_{\mu\nu}^{(iv)-(vii)} = \frac{1}{2N} \sum_n \sum_{\tilde{n}(>n)}\sum_\mathbf{k} \langle u_{{n}}(\mathbf{k})|\tau_z |u_n(\mathbf{k})  \rangle&\bigg( \langle \partial_{k_\nu} u_n(\mathbf{k})|u_{\tilde{n}}(\mathbf{k}) \rangle\langle u_{\tilde{n}}(\mathbf{k})|\tau_z |u_n(\mathbf{k})  \rangle + \langle u_{\tilde{n}}(\mathbf{k}) |\partial_{k_\nu} u_n(\mathbf{k}) \rangle\langle u_{{n}}(\mathbf{k})|\tau_z |u_{\tilde{n}}(\mathbf{k})  \rangle \bigg) \nonumber \\ 
    \times \partial_{k_\mu} &\bigg[ 2\frac{f( E_{\tilde{n}}(\mathbf{k})) - f( E_{{n}}(\mathbf{k}))}{E_n(\mathbf{k}) - E_{\tilde{n}}(\mathbf{k})} + f'(E_n(\mathbf{k})) + f'(E_{\tilde{n}}(\mathbf{k}))\bigg] + \mu \leftrightarrow \nu \nonumber \\
     =-\frac{1}{2} \int\frac{d\mathbf{k}}{(2\pi)^d}\sum_n \sum_{\tilde{n}(> n)} \partial_{k_\mu} &\bigg( \langle u_{{n}}(\mathbf{k})|\tau_z |u_n(\mathbf{k})  \rangle \langle \partial_{k_\nu} u_n(\mathbf{k})|u_{\tilde{n}}(\mathbf{k}) \rangle\langle u_{\tilde{n}}(\mathbf{k})|\tau_z |u_n(\mathbf{k})  \rangle + c.c. \bigg) \nonumber \\ 
    \times  &\bigg[ f'(E_n(\mathbf{k})) + f'(E_{\tilde{n}}(\mathbf{k})) + 2\frac{f(E_{\tilde{n}}(\mathbf{k})) - f(E_{n}(\mathbf{k}))}{E_n(\mathbf{k}) - E_{\tilde{n}}(\mathbf{k})}\bigg] + \mu \leftrightarrow \nu 
\end{align}
where we integrated by parts in the last step. Finally, we can collect all terms
\begin{align}\label{eq:all_terms_general_final}
    \lim_{\mathbf{q}\rightarrow 0} \partial_{q_\mu}\partial_{q_\nu}\chi(0, \mathbf{q}) = \sum_{n}\int\frac{d\mathbf{k}}{(2\pi)^d}&\partial_{k_\mu}E_n(\mathbf{k})\partial_{k_\nu}E_n(\mathbf{k}) \frac{f^{(3)}(E_n(\mathbf{k}))}{6}\left|\langle u_{n}(\mathbf{k})|\tau_z|u_{{n}}(\mathbf{k})\rangle\right|^2 \nonumber \\
    +\sum_{n}\sum_{\tilde{n}(>n)}\int\frac{d\mathbf{k}}{(2\pi)^d} &\bigg( f'(E_n(\mathbf{k})) + f'(E_{\tilde{n}}(\mathbf{k})) + 2\frac{f(E_{\tilde{n}}(\mathbf{k})) - f(E_{n}(\mathbf{k}))}{E_n(\mathbf{k}) - E_{\tilde{n}}(\mathbf{k})}\bigg)\nonumber \\ \times \bigg[ &\frac{v_{n,\mu}(\mathbf{k})v_{\tilde{n},\nu}(\mathbf{k})\left| \langle u_{{n}}(\mathbf{k})|\tau_z |u_{\tilde{n}}(\mathbf{k})  \rangle  \right|^2 
    }{(E_n(\mathbf{k}) - E_{\tilde{n}}(\mathbf{k}))^2} + A_{n\tilde{n},\mu}(\mathbf{k}) A_{\tilde{n}n,\nu}(\mathbf{k}) \langle u_{n}(\mathbf{k})|\tau_z|u_{{n}}(\mathbf{k})\rangle\langle u_{\tilde{n}}(\mathbf{k})|\tau_z|u_{\tilde{n}}(\mathbf{k})\rangle \nonumber \\ -\frac{1}{2} \partial_{k_\mu}&\bigg( \langle u_{{n}}(\mathbf{k})|\tau_z |u_n(\mathbf{k})  \rangle \langle \partial_{k_\nu} u_n(\mathbf{k})|u_{\tilde{n}}(\mathbf{k}) \rangle\langle u_{\tilde{n}}(\mathbf{k})|\tau_z |u_n(\mathbf{k})  \rangle + c.c.\bigg) \bigg] + \mu \leftrightarrow \nu,
\end{align}
where we defined the (interband) Berry connection as $A_{n\tilde{n},\mu}(\mathbf{k}) = i\langle \partial_{k_\mu}u_n(\mathbf{k})|u_{\tilde{n}}(\mathbf{k})\rangle$. 

In the next section, we apply these formulae to a specific class of effective models and will assume a specific type of fluctuation (that is, altermagnetic fluctuations with $\tau_\mathrm{M}=\tau_z$). Even without making these assumptions, one can identify geometric quantities like the quantum metric, the positional shift or the interband dipole copuling in the expression above. For an arbitrary number of bands, additional terms like $n\neq n' \neq \tilde{n} \neq \tilde{n}'$ etc. have to be considered. Working out a general expression for any type of correlation function in a generic (non-interacting) multi-band system, and relating it to quantum geometry is left for future work. 

\subsection{Altermagnetic Fluctuations}

We now focus on the altermagnetic fluctuations with $\tau_\mathrm{M}=\tau_z$ \mbox{(see Eq.\ \ref{eq:greens_function_tau_M})} of an effective altermagnetic two-band model of the form
\begin{align}\label{eq:2band_ham}
    \mathcal{H}_0(\mathbf{k}) = \varepsilon_0(\mathbf{k})\mathbbm{1}_{2\times 2} + t_x(\mathbf{k})\tau_x + t_z(\mathbf{k})\tau_z.
\end{align}
As argued in the main text, these models are shown to capture many properties of a range of altermagnetic material candidates, and have been shown to have leading altermagnetic instabilities for finite $t_x$ and $t_z$\cite{roig2024minimal}. Our goal is to understand the nature of these instabilities and their relation to quantum geometry by identifying geometric quantities in Eq.\ \ref{eq:all_terms_general_final}. Assuming a specific model is not necessary to find these quantities in what we derived, but it makes notation much cleaner. Moreover -- because $\mathcal{H}_0$ commutes with $\tau_y$ -- the Hamiltonian is real and we can express its eigenstates as unit vectors on the Bloch sphere at fixed azimuthal angle given by\cite{roig2024minimal}
\begin{align}\label{eq:eigenvectors_model}
    U(\mathbf{k}) = \begin{pmatrix}
    \cos \frac{\theta(\mathbf{k})}{2} & \sin \frac{\theta(\mathbf{k})}{2}\\
    -\sin \frac{\theta(\mathbf{k})}{2} & \cos \frac{\theta(\mathbf{k})}{2}
    \end{pmatrix}
\end{align}
with 
\begin{align}\label{eq:sin_theta_cos_theta}
    \sin \theta(\mathbf{k}) = \frac{t_x(\mathbf{k})}{\sqrt{t_x^2(\mathbf{k}) + t^2_z(\mathbf{k})}} \qquad \mathrm{and} \qquad \cos \theta(\mathbf{k}) = \frac{t_z(\mathbf{k})}{\sqrt{t_x^2(\mathbf{k}) + t^2_z(\mathbf{k})}}.
\end{align}
We note that $\tau_z$ is diagonal in the orbital basis with $\tau_\mathrm{z}|1\rangle = +|1\rangle$ and $\tau_\mathrm{z}|2\rangle = -|2\rangle$ with 1 and 2 denoting the two sublattices of the minimal altermagnetic model. 
The matrix elements of $\tau_z$ in the band basis are then given by
\begin{align}
    \langle u_{n}(\mathbf{k})|\tau_z|u_{\tilde{n}}(\mathbf{k})\rangle &= \langle u_{n}(\mathbf{k})|{\bigg(} |1\rangle \langle 1| + |2\rangle \langle 2|{\bigg)}\tau_z {\bigg(} |1\rangle \langle 1| + |2\rangle \langle 2|{\bigg)}|u_{\tilde{n}}(\mathbf{k})\rangle \nonumber \\ 
    &= \langle u_{n}(\mathbf{k})|1\rangle\langle 1| \tau_z |1\rangle\langle 1|u_{\tilde{n}}(\mathbf{k})\rangle + \langle u_{n}(\mathbf{k})|2\rangle\langle 2| \tau_z |2\rangle\langle 2|u_{\tilde{n}}(\mathbf{k})\rangle \nonumber \\
    &= \langle u_{n}(\mathbf{k})|1\rangle\langle 1|u_{\tilde{n}}(\mathbf{k})\rangle - \langle u_{n}(\mathbf{k})|2\rangle\langle 2|u_{\tilde{n}}(\mathbf{k})\rangle. 
\end{align}
We know $U$, so we can explicitly write down the matrix as
\begin{align}\label{eq:overlap_matrix}
    \left[\langle u(\mathbf{k})|\tau_z|u(\mathbf{k})\rangle\right]_{n\tilde{n}} = \begin{pmatrix}
    \cos^2 \frac{\theta(\mathbf{k})}{2} & \frac{1}{2}\sin \theta(\mathbf{k})\\
    \frac{1}{2}\sin \theta(\mathbf{k}) & \sin^2 \frac{\theta(\mathbf{k})}{2}
    \end{pmatrix} - \begin{pmatrix}
    \sin^2 \frac{\theta(\mathbf{k})}{2} & -\frac{1}{2}\sin \theta(\mathbf{k})\\
    -\frac{1}{2}\sin \theta(\mathbf{k}) & \cos^2 \frac{\theta(\mathbf{k})}{2}
    \end{pmatrix} = \begin{pmatrix}
    \cos \theta(\mathbf{k}) & \sin \theta(\mathbf{k})\\
    \sin \theta(\mathbf{k}) & -\cos \theta(\mathbf{k})
    \end{pmatrix},
\end{align}
where we used the trigonometric identity
\begin{align}\label{eq:trig_id}
    \cos^2 x - \sin^2 x = \cos 2x.
\end{align}
We note that this family of models naturally fulfill the condition that we used in the previous section to rewrite Eq.\ \ref{eq:chi_fourseven_neqntilde} as Eq.\ \ref{eq:chi_fourseven_before_product_rule}, namely that 
\begin{align}\label{eq:diagonal_odd}
    \langle u_1(\mathbf{k})|\tau_z|u_1(\mathbf{k})\rangle = -\langle u_2(\mathbf{k})|\tau_z|u_2(\mathbf{k})\rangle
\end{align}
Knowing the explicit form of eigenstates \mbox{(see Eq.\ \ref{eq:eigenvectors_model})}, 
we rewrite $\langle \partial_{k_\nu} u_n(\mathbf{k})|u_{\tilde{n}}(\mathbf{k}) \rangle$ as the rate of change of the (elevation) Bloch angle 
\begin{align}
    \langle \partial_{k_\nu} u_1(\mathbf{k})|u_{2}(\mathbf{k}) \rangle = \left(\partial_{k_\nu} \cos \frac{\theta(\mathbf{k})}{2} \ -\partial_{k_\nu}\sin \frac{\theta(\mathbf{k})}{2}\right)\begin{pmatrix}
        \sin \frac{\theta(\mathbf{k})}{2} \\
        \cos \frac{\theta(\mathbf{k})}{2}
    \end{pmatrix}= - \frac{\theta^\nu(\mathbf{k})}{2}\sin^2\frac{\theta(\mathbf{k})}{2} - \frac{\theta^\nu(\mathbf{k})}{2}\cos^2 \frac{\theta(\mathbf{k})}{2} = -\frac{\theta^\nu(\mathbf{k})}{2},
\end{align}
where we denote the derivative of $\theta$ w.r.t $k_\nu$ as $\theta^\nu$. Inserting the explicit form of the elements of $\tau_z$ \mbox{(see Eq.\ \ref{eq:overlap_matrix})}, we rewrite the partial derivative in the last line of Eq.\ \ref{eq:all_terms_general_final} as
\begin{align}\label{eq:metric_in_angle}
    \partial_{k_\mu}\left(\langle u_{{1}}(\mathbf{k})|\tau_z |u_1(\mathbf{k})  \rangle \langle \partial_{k_\nu} u_1(\mathbf{k})|u_{2}(\mathbf{k}) \rangle\langle u_{2}(\mathbf{k})|\tau_z |u_1(\mathbf{k})  \rangle + c.c.\right) &= -\partial_{k_\mu}\left(\cos \theta(\mathbf{k}) \sin \theta(\mathbf{k}) \theta^\nu(\mathbf{k})\right)\nonumber \\
    &= -\frac{1}{2}\partial_{k_\mu}\left( \sin 2\theta(\mathbf{k}) \theta^\nu(\mathbf{k})\right)\nonumber \\
    &= -\cos 2\theta(\mathbf{k}) \theta^\mu(\mathbf{k})\theta^\nu(\mathbf{k})- \frac{1}{2} \sin 2\theta(\mathbf{k}) \partial_{k_\mu}\theta^\nu(\mathbf{k}).
\end{align}
For a general multi-band system, the quantum metric is defined as
\begin{align}\label{eq:metric_connection}
    g_{n,\mu\nu}(\mathbf{k}) =\sum_{\tilde{n}(\neq n)} A_{n\tilde{n},\mu}(\mathbf{k})A_{\tilde{n}n,\nu}(\mathbf{k}) + c.c.,
\end{align}
and specifically for our two-band model 
\begin{align}\label{eq:metric_matel}
    g_{1,\mu\nu}(\mathbf{k}) = -\langle \partial_{k_\mu}u_{1}(\mathbf{k})|u_{2}(\mathbf{k})\rangle \langle \partial_{k_\nu}u_{2}(\mathbf{k})|u_{1}(\mathbf{k})\rangle + c.c. =  \frac{1}{2}\theta^\mu(\mathbf{k})\theta^\nu(\mathbf{k}).
\end{align}
Dropping the band index we denote the metric as $g_{\mu\nu}$ and since we are interested in its diagonal elements, we focus on the case $\mu=\nu$. Inserting the above expressions into Eq.\ \ref{eq:all_terms_general_final} we find
\begin{align}\label{eq:fluct_sec_final_expr}
    \lim_{\mathbf{q}\rightarrow 0} \partial_{q_\mu}\partial_{q_\mu}\chi(0, \mathbf{q}) =  \int\frac{d\mathbf{k}}{(2\pi)^d}\frac{1}{6}\left( f^{(3)}(E_1(\mathbf{k}))v^2_{1,\mu}(\mathbf{k}) + f^{(3)}(E_2(\mathbf{k}))v^2_{2,\mu}(\mathbf{k})\right)\cos^2\theta(\mathbf{k})\nonumber \\
    +\int\frac{d\mathbf{k}}{(2\pi)^d} \bigg(2\frac{v_{1,\mu}(\mathbf{k})v_{2,\mu}(\mathbf{k})}{(E_n(\mathbf{k}) - E_{\tilde{n}}(\mathbf{k}))^2}\sin^2\theta(\mathbf{k}) + g_{\mu\mu} \left(5\cos^2\theta(\mathbf{k}) - 2\right) + \frac{1}{2}\sin 2\theta(\mathbf{k}) \partial_{k_\mu}\theta^\mu(\mathbf{k})
    \bigg)  \nonumber \\ \times \bigg( {f'(E_1(\mathbf{k})) + f'(E_{2}(\mathbf{k}))} + 2\frac{f(E_{2}(\mathbf{k})) - f(E_{1}(\mathbf{k}))}{E_1(\mathbf{k}) - E_{2}(\mathbf{k})}\bigg) 
\end{align}
where we used the trigonometric identity
\begin{align}
    2\cos(2x) + \cos^2(x) = 5\cos^2(x) - 2.
\end{align}
Applying the product rule to second derivative of the Bloch angle in Eq.\ \ref{eq:fluct_sec_final_expr} yields
\begin{align}
    -\partial_\mu\frac{\theta^\nu(\mathbf{k})}{2}\partial_{k_\mu} \langle \partial_{k_\nu} u_1(\mathbf{k})|u_{2}(\mathbf{k}) \rangle = \langle \partial_{k_\mu}\partial_{k_\nu} u_1(\mathbf{k})|u_{2}(\mathbf{k}) \rangle + \langle \partial_{k_\nu} u_1(\mathbf{k})|\partial_{k_\mu}u_{2}(\mathbf{k}) \rangle.
\end{align}
An interpretation of this tensor is provided in Ref.\ \cite{cheng2013quantumgeometrictensorfubinistudy}, where the gauge-dependent part of the quantum metric $\gamma_{\mu\nu}$ is introduced. It measures the distance of bare states in the Hilbert space, while the distance of rays in the projected (or normalized) Hilbert space is given by the quantum metric $g_{\mu\nu}$. It is shown that $\gamma_{\mu\nu} = \mathrm{Re} \langle \partial_{k_\mu} u(\mathbf{k})|\partial_{k_\nu}u(\mathbf{k}) \rangle = -\mathrm{Re}\langle u(\mathbf{k})|\partial_{k_\mu}\partial_{k_\nu}u(\mathbf{k}) \rangle$, and because our Hamiltonian is real, this can be discarded. 

In analogy to Ref.\ \cite{taisei}, we define $\chi^{0:\mu\nu}_\mathrm{m} = \chi^{0:\mu\nu}_\mathrm{m:mass} + \chi^{0:\mu\nu}_\mathrm{m:velo} + \chi^{0:\mu\nu}_\mathrm{m:geom}$ with
\begin{align}
    \chi^{0:\mu\nu}_\mathrm{m:mass} = \frac{1}{6}\int\frac{d\mathbf{k}}{(2\pi)^d}\left( f^{(3)}(E_1(\mathbf{k}))v^2_{1,\mu}(\mathbf{k}) + f^{(3)}(E_2(\mathbf{k}))v^2_{2,\mu}(\mathbf{k})\right)\cos^2\theta(\mathbf{k})
\end{align}
\begin{align}
    \chi^{0:\mu\nu}_\mathrm{m:velo} = 2\int\frac{d\mathbf{k}}{(2\pi)^d} \frac{v_{1,\mu}(\mathbf{k})v_{2,\mu}(\mathbf{k})}{(E_1(\mathbf{k}) - E_{2}(\mathbf{k}))^2}\left( f'(E_1(\mathbf{k})) + f'(E_{2}(\mathbf{k})) + 2\frac{f(E_{2}(\mathbf{k})) - f(E_{1}(\mathbf{k}))}{E_1(\mathbf{k}) - E_{2}(\mathbf{k})}\right) \sin^2\theta(\mathbf{k})
\end{align}
\begin{align}
    \chi^{0:\mu\nu}_\mathrm{m:geom} = \int\frac{d\mathbf{k}}{(2\pi)^d} g_{\mu\mu} \left( f'(E_1(\mathbf{k})) + f'(E_{2}(\mathbf{k})) + 2\frac{f(E_{2}(\mathbf{k})) - f(E_{1}(\mathbf{k}))}{E_1(\mathbf{k}) - E_{2}(\mathbf{k})}\right) \left(5\cos^2\theta(\mathbf{k}) - 2\right).
\end{align}

\subsection{Ab-initio and Effective model of MnTe}
\begin{figure}
    \centering
    \includegraphics[width=0.8\linewidth]{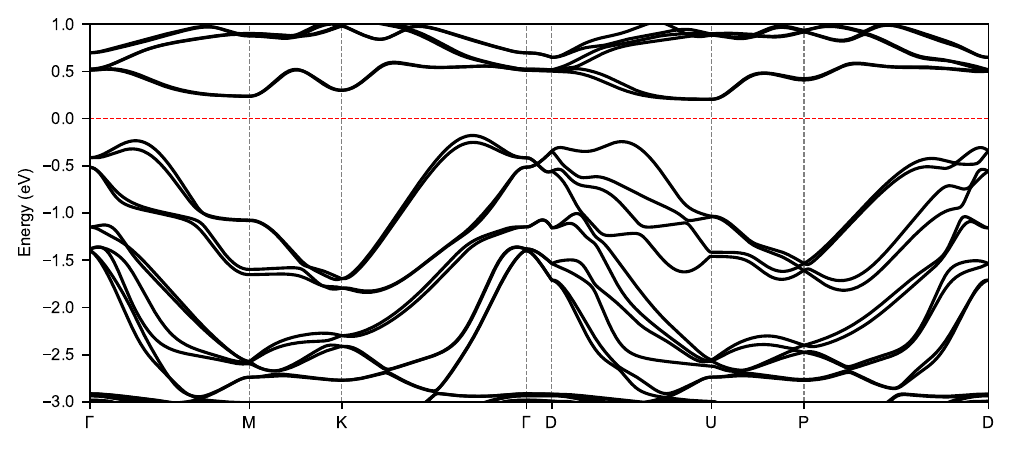}
    \caption{Electronic band structure of MnTe in its N\'eel-ordered phase with $3.88\mu_{\mathrm{B}}$ on the Mn. Along D-U there is a notable spin-splitting of approximately $0.4$ eV.}
    \label{fig:dft}
\end{figure}

We describe the normal state band structure of MnTe using a real-space, two-site tight-binding model of the form
\begin{align}
    \mathcal{H}_0 = \sum_{\langle ij \rangle_\alpha} t_{\alpha}(c_i^\dagger c_j + c.c.) - \mu + \sum_i \left(-1\right)^i\Delta\sigma_z
\end{align}
that we fit with $\Delta=0.2$ eV to the four highest occupied bands of altermagnetic MnTe as obtained from ab-initio on a $12^3$-grid using the generalized gradient approximation (GGA)\cite{perdew1996generalized} as implemented in the VASP package\cite{vasp1,vasp2,vasp3}. We show the band structure along $\Gamma$-M-K-$\Gamma$-D-U-P-D in Fig.\ \ref{fig:dft}. 
We choose the lattice constants as $a=b=4.17349018$ and $c=6.75345133$. The Mn atoms sit on Wyckoff position $2a$ at $(0, 0, 0)$ and $(0, 0, 0.5)$. $\alpha$ iterates over the (symmetry-inequivalent) neighbors with the fitted hopping amplitudes $t_\alpha$ given in Tab.\ \ref{tab:parameters}. The chemical potential is $\mu = 1.07250035$ eV. The tenth- and eleventh-nearest neighbors (at a distance of 9.89258 \AA) are the first hoppings that result in finite spin-splitting and manifest the true space group symmetry of the full lattice. They have the same distance, but cannot be mapped onto each other by any symmetry in $P6_3/mmc$. Rather than fitting the normal state effective model to the spin-polarized ab-initio results, we use the N\'eel-ordered band structure because density functional theory of non-magnetic MnTe falsely predicts it to be metallic (see supplemental material of \cite{mnte_prl}). The magnetic relaxation yields the correct anti-ferromagnetic configuration (without any $+U$) with a magnetic moment of $\pm3.88\mu_{\mathrm{B}}$ on the Mn sites. The magnetic band structure is insulating and shows the expected altermagnetic spin-splitting along the off-nodal ($k_z=0.3\pi/c$) path segment D-U.

\begin{table}[]
    \centering
    \begin{tabular}{c|c|c|c|c}
        $\alpha$ & distance (\AA) & coordination number & type & $t_\alpha$ (eV)  \\ \hline 
        1 & 3.37673 & 2 & $t_x$ & 0.05900989 \\
        2 & 4.17349 & 6 & $\varepsilon_0$ & 0.0805197 \\
        3 & 5.36845 & 12 & $t_x$ & 0.01127167 \\
        4 & 6.75345 & 2 & $\varepsilon_0$ & -0.03001691 \\
        5 & 7.2287 & 6 & $\varepsilon_0$ & -0.00338583 \\
        6 & 7.93896 & 12 & $\varepsilon_0$ & 0.02484375 \\
         7 & 7.97849 & 12 &$t_x$ & -0.00427934 \\
        8 & 8.34698 & 6 & $\varepsilon_0$ & 0.02986773 \\
        9 & 9.00413 & 12 & $t_x$ & -0.01658517 \\
        10 & 9.89258 & 6 & $t_z$ & -0.05772139 \\
        11 & 9.89258 & 6 & $t_z$ & 0.02127313 
    \end{tabular}
    \caption{Parameters of the tight-binding model of MnTe.}
    \label{tab:parameters}
\end{table}

\end{document}